\documentclass[pr,twocolumn,preprintnumbers,nofootinbib,showpacs,amsmath,amssymb,superscriptaddress]{revtex4-1}
\usepackage{amsfonts}
\usepackage{mathrsfs}
\usepackage{graphicx}
\usepackage{amssymb}
\usepackage{color}
\usepackage{amsmath}
\usepackage{CJK}
\usepackage[
      colorlinks=true,
      linkcolor=blue,
      urlcolor=blue,
      filecolor=black,
      citecolor=blue,
            ]{hyperref}

\usepackage{slashed}
\usepackage{epstopdf}

\begin{document}

\title{ Massive $2$-form field and holographic ferromagnetic phase transition}
\author{Rong-Gen Cai}
\email{cairg@itp.ac.cn}

\author{Run-Qiu Yang}
\email{aqiu@itp.ac.cn}
\affiliation{State Key Laboratory of Theoretical Physics,Institute of Theoretical Physics,\\
 Chinese Academy of Sciences,Beijing 100190, China.}

\author{Ya-Bo Wu}
\author{Cheng-Yuan Zhang}
\affiliation{Department of Physics, Liaoning Normal University, Dalian, 116029, China}

\begin{abstract}
In this paper we investigate in some detail the holographic ferromagnetic phase transition in an AdS${_4}$  black brane background by introducing a massive 2-form field coupled to the
Maxwell field strength in the bulk.  In the two probe limits, one is to neglect the back reaction of the 2-form field to the background geometry and to the Maxwell field, and the other to neglect
the back reaction of  both the Maxwell field and the 2-form field, we find that  the spontaneous magnetization and the ferromagnetic phase transition always happen when the temperature
gets low enough with similar critical behavior. We calculate  the DC resistivity in a semi-analytical method in the second probe limit  and find  it behaves as the colossal magnetic resistance effect in
some materials.  In the case with the first probe limit, we obtain the off-shell free energy of the holographic model near the critical temperature and compare with  the Ising-like model.  We also study the
back reaction effect and find that the phase transition is always second order.  In addition, we find  an analytical Reissner-Norstr\"om-like black brane solution in the Einstein-Maxwell-2-form field theory
with a negative cosmological constant.

\end{abstract}
\maketitle
\tableofcontents

\section{Introduction}

In recent years, a promising new route named  AdS/CFT duality~\cite{1,2,3,4} provided a new viewpoint to understand  gravity and strongly coupled or correlated phenomena in physics. This duality relates a gravity theory in a weakly curved ($d+1$)-dimensional anti de Sitter (AdS$_{d+1}$)
spacetime to a strongly-coupled $d$-dimensional field theory living on its boundary. AdS/CFT maps questions about strongly correlated many-body phenomena to solvable single- or few-body classical problems in a curved geometry, which opens a new window to solve the strongly correlated system in condensed matter physics and considerable progresses have been made since the duality
was proposed. For instance, holographic superconductor/superfuild~\cite{Hartnoll:2008vx,Hartnoll:2008kx} (for reviews, see \cite{Hartnoll:2009sz,Herzog:2009xv,McGreevy:2009xe,Horowitz:2010gk,Cai:2015cya}), holographic (non-)fermi liquid~\cite{Lee:2008xf,Liu:2009dm,Cubrovic:2009ye}, holographic charge density wave and metal/insulator phase transition~\cite{Aperis:2010cd,Donos:2013gda,Ling:2014saa},  and some systems far-from thermal equilibrium~\cite{Murata:2010dx,Bhaseen:2012gg,Adams:2012pj,Garcia-Garcia:2013rha,Chesler:2013lia} have been studied intensively.

So far, most attentions about the duality application to condensed matter physics have been focused on the electronic properties of materials.  In condensed matter physics, magnetism also plays an important role in materials including high temperature superconductors and heavy fermion metals in many strongly correlated electronic systems. The gauge/gravity duality provides  an approach
and perspective to understand these challenging propblems. Though there exist a few works  involving magnetism in  holographic superconductor models, magnetism does not pay the central role. Yet the scarcity of models on magnetism is  due to various technical challenges in holographic context. In a previous work~\cite{29}, we first proposed a new example of the application of the AdS/CFT correspondence by realizing the paramagnetism/ferromagnetism phase transition in a dyonic Reissner-Nordatr\"{o}m-AdS black brane. This model also was extended to realize the paramagnetism/antiferromangnetism phase transition by introducing two real antisymmetric tensor fields with interaction between them and they are coupled to the background gauge field strength~\cite{Cai:2014jta}. In Ref.~\cite{31},
by combining the holographic $p$-wave superconductor model \cite{Cai:2013aca} and the holographic ferromagnetism model,
we studied the coexistence and competition of ferromagnetism and $p$-wave superconductivity.
It is found that the results depend on the self-interaction of magnetic moment of the complex vector field and which phase (superconductivity or ferromagnetism) appears first. We noted that the ferromagnetic superconductivity has been also discussed in Ref.~\cite{Amoretti:2013oia} by introducing two SU(2) gauge fields in the bulk. In that model superconductivity and ferromagnetism happens simultaneously and the magnetic susceptibility is finite at the ferromagnetic critical temperature. However, the ferromagnetic $p$-wave superconductor in heavy fermion systems such as UCoGe, URhGe and UGe$_2$, shows the two critical temperatures are different in general~\cite{F.Hardy} and magnetic susceptibility diverges at ferromagnetic critical temperature~\cite{N.Tateiwa}.

The reasons of using an antisymmetric tensor field to model ferromagnetic phase transition in holographic setup are as follows.
On the viewpoint  of symmetry breaking, ferromagnetic phase transition breaks the time reversal symmetry spontaneously in low temperature
(if spatial dimension is more than 2, it also breaks spatial rotation symmetry),
which in general is not associated with other  symmetry breakings such as U(1), SU(2) and so on.
Thus the dual operator does not carry U(1) or SU(2) charge, which implies that we need to use a real field corresponding to such an operator.
From  the point of view of covariance, magnetic field is the spatial component of a SO(1,3) tensor $F_{\mu\nu}$, so magnetic moment should be also a component of a tensor.
Furthermore, according to the origin of magnetic moment, magnetism of material is controlled by the intrinsic magnetism of electrons which is obtained from the Lagrangian for electrons,
\begin{equation}\label{Diraca1}
\mathcal{L}_e=i(\overline{\psi}\slashed\partial\psi-m\overline{\psi}\psi)-ej^\mu A_\mu-\frac{e}{4m}\overline{\psi}\sigma^{\mu\nu}\psi F_{\mu\nu}.
\end{equation}
where $A_\mu$ is the gauge potential, $F_{\mu\nu}$ is the gauge strength field and $j^\mu$ is current density which is defined as $j^\mu=-\frac{i}{2m}\overline{\psi}(\overrightarrow{\partial}^\mu-\overleftarrow{\partial}^\mu)\psi$. Then the magnetic moment density $\overrightarrow{N}$ is the spatial component of an antisymmetric tensor field,
\begin{equation}\label{MNDirac}
    N^i={\epsilon^i}_{jk}M^{jk},~M^{\mu\nu}=\frac{e}{4m} \langle \overline{\psi}\sigma^{\mu\nu}\psi \rangle.
\end{equation}
Here $i,j,k=1,2,3$. Thus we see that an effective field to describe magnetic moment in the boundary field in a covariant manner needs an antisymmetric tensor and its spatial component corresponds to the magnetic moment. These considerations can be generalized into $2+1$ dimensions, where magnetic moment should still be regarded as the spatial components of an antisymmetric tensor field. Then according to the spirit of AdS/CFT, we can use an antisymmetric tensor field in the bulk to dual such a tensor operator.

Recently,  it has been found in Ref.~\cite{Cai:2015bsa} that the original model in \cite{29}  contains a ghost and a new model has been proposed, there the exterior differential is used to replace
 the covariant derivative in the kinetic term of the antisymmetric tensor field. With such a simple modification it has been shown that there does not exist any ghost and causality violation
does not appear in the new model, while the key results in the original model are kept qualitatively.  While in Ref.~\cite{Cai:2015bsa}  we paid our main attention to the health of the model and
showed that the spontaneous magnetization can happen when the temperature gets low enough, in this paper, we are going to study this model in some detail with some directions.
Before going on, we mention here that in this paper we  assume the physical phase is homogeneous and will not consider inhomogeneous phases. This assumption is just for simplifying our discussions in technology. The possibility whether the inhomogeneous phase could appear spontaneously in this model under the homogeneous boundary conditions is left to study in the future.

This paper is organized as follows.
In section \ref{model}, we will describe the holographic model.
We will give our ansatz for matter field and derive equations of motion in section \ref{sect3}. In this section, we will show that there exists an analytical black hole solution different from the AdS Reissner-Nordtrs\"om  solution in this model and discuss associated properties of the solution. In sections \ref{prob1} and \ref{prob2}, we will investigate the paramagnetism/ferromagnetism phase transition
in two different probe limits.  One is to neglect the back reaction of the 2-form field to the black brane geometry and to the Maxwell field, and the other to neglect the back reaction of both the Maxwell field and the form field.  In the
former case, we will also calculate the off-shell free energy of the holographic model near the phase transition point by using the Sturm-Liouville eigenvalue method and obtain a  free energy form like
the Ginzburg-Landau one, and make a comparison with the Ising like university class model.  In the latter case, we calculate the DC resistivity in the ferromagnetic phase, which  shows the behavior of the colossal magnetic resistance effect in condensed matter physics. We study the full back reaction effect in  section~\ref{backrec} by solving the full equations of motion of the model and find that the phase transition is aways second order. The summary and some discussions are included in section \ref{summ}.

\section{The Model}
\label{model}
In this paper, the model we are considering is just general relativity with
a negative cosmological constant $\Lambda=-3/L^2$,
 a U(1) field $A_{\mu}$ and a massive 2-form field $M_{\mu\nu}$ in
4-dimension space-time. The ghost free action reads~\cite{Cai:2015bsa}
\begin{eqnarray}\label{1}
S&=&\frac{1}{2\kappa^2}\int d^4x\sqrt{-g}(L_{1}+\lambda^2 L_{2}),\nonumber\\
L_{1}&=&\mathcal{R}+\frac{6}{L^2}-F^{\mu\nu}F_{\mu\nu},\\
L_{2}&=&-\frac{1}{12}(d M)^2-\frac{m^2}{4}M_{\mu\nu}M^{\mu\nu}-M^{\mu\nu}F_{\mu\nu}-\frac{J}{8}V(M).\nonumber
\end{eqnarray}
where $L$ is the AdS radius which is set to be unity
and $2\kappa^2=16\pi G$ is associated to the gravitational constant in the bulk, which will be set to be unity in the following.
$g$ is the determinant of the bulk metric $g_{\mu\nu}$.
$d M$ is the exterior differential of 2-form field $M_{\mu\nu}$.
$m^2\neq0$ is the squared mass of 2-form field $M_{\mu\nu}$
and must be greater than zero, which will be explained shortly.
$\lambda$ and $J$ are two real model parameters with $J<0$ in order the magnetization to happen spontaneously.
$\lambda^2$ characterizes the back reaction of the 2-form field $M_{\mu\nu}$ to the background geometry and to the Maxwell field strenghth~\footnote{Note that here $\lambda$ can be also
understood as the measurement of the coupling between the tensor field $M_{\mu\nu}$ and the Maxwell field strength  $F_{\mu\nu}$ by rescaling the tensor field and the parameter $J$. }.
$V(M_{\mu\nu})$ is a nonlinear potential of the 2-form field.
It describes the self-interaction of the polarization tensor,
which should be expanded as the even power of $M_{\mu\nu}$.
In this model, we take the following form,
\begin{eqnarray}\label{2}
V(M_{\mu\nu})&=&(^{*}M_{\mu\nu}M^{\mu\nu})^2=[^{*}(M\wedge M)]^2.
\end{eqnarray}
Here $^{*}$ is the Hodge-star operator. The choice of nonlinear potential is not unique.
We choose this form just for simplicity. As shown in Ref.~\cite{Cai:2015bsa}, this potential shows a global minimum at
some nonzero value of $\rho$.

By varying  action \eqref{1}, we can get the equations of motion  for the matter fields and gravitational field as
\begin{eqnarray}\label{3}
\nabla^{\tau}(d M)_{\tau\mu\nu}-m^2 M_{\mu\nu}-J(^{*}M_{\tau\sigma}M^{\tau\sigma})(^{*}M_{\mu\nu})&=&F_{\mu\nu},\nonumber\\
\nabla^{\mu}(F_{\mu\nu}+\frac{\lambda^2}{4}M_{\mu\nu})&=&0,\\
R_{\mu\nu}-\frac{1}{2} R g_{\mu\nu}-\frac{3}{L^2}g_{\mu\nu}&=&T_{\mu\nu}.\nonumber
\end{eqnarray}
The energy-momentum tensor $T_{\mu\nu}$ reads
\begin{eqnarray}\label{4}
T_{\mu\nu}&=&\lambda^2[\frac{1}{4}(d M)_{\sigma\mu\tau} (d M)^{\sigma\alpha\tau} g_{\alpha\nu}+\frac{m^2}{2}M_{\tau\mu}{M^{\tau}}_{\nu}\nonumber\\
&&{}+M_{\tau(\nu} {F^{\tau}}_{\nu)}+\frac{J}{8}V(M) g_{\mu\nu}]+2{F^{\tau}}_{\mu}F_{\tau\nu}\nonumber\\
&&+\frac{1}{2}(\lambda^2L_{2}-F_{\mu\nu}F^{\mu\nu})g_{\mu\nu}.
\end{eqnarray}

In the AdS/CFT correspondence, a hairy black hole with appropriate boundary conditions
can be explained as a condensed phase of the dual field theory,
while a black hole without hair is dual to a normal phase. Clearly when $M_{\mu\nu}$ vanishes, the model admits
the AdS Reissner-Nordstr\"om (RN) black brane solution, which corresponds to  the normal phase in the dual field theory. When $M_{\mu\nu}$ appears,
we will see that there exists an analytical black brane solution differing from the AdS-RN solution.  The new solution corresponds to the normal phase rather than
the usual AdS RN solution.  When we lower the temperature, the system exhibits an instability which triggers to break time reversal symmetry spontaneously
as well as spatial rotation symmetry since the condensate will pick out one direction as special
(if spatial dimension is more than $2$) and the paramagnetism/ferromagnetism phase transition happens.

\section{Ansatz and trivial solution}
\label{sect3}
As we will discuss the full solution to the action \eqref{1} including
the back reaction to the spacetime geometry, we start with the following ansatz for the metric
\begin{eqnarray}\label{5}
ds^2&=&-r^2 f(r) e^{a(r)}dt^2+\frac{dr^2}{r^2 f(r)}+r^2(dx^2+dy^2),
\end{eqnarray}
and take the self-consistent ansatz for polarization field and U($1$) field as
\begin{eqnarray}\label{6}
M_{\mu\nu}&=&-p(r)dt\wedge dr+\rho(r)dx\wedge dy,\nonumber\\
A_{\mu}&=&\phi(r)dt+Bx dy,
\end{eqnarray}
with some real functions $f(r)$, $a(r)$, $\phi(r)$, $p(r)$,and $\rho(r)$.
The bulk field $B$ is a constant magnetic field,
which can be regarded as external magnetic field in the dual boundary theory.
We will denote the position of the horizon as $r_{h}$ and the conformal boundary will
be at $r\rightarrow\infty$.
Since we would like to study a dual theory with finite chemical potential or
charge density accompanied by a U(1) symmetry, we turn on $A_{t}$ in the bulk.

Suppose the black brane horizon is at $r_h$ with $f(r_{h})=0$.
According to gauge/gravity duality, the Hawking temperature of the black brane is
identified with the temperature of boundary thermal state, which is given by
\begin{eqnarray}\label{7}
T=\left.\frac{(r^{2} f(r))' e^{a(r)/2}}{4 \pi}\right|_{r=r_h},
\end{eqnarray}
and the thermal entropy $S$ is given by the Bekenstein-Hawking entropy of the black brane
\begin{eqnarray}\label{8}
S&=&4\pi A=4\pi r_{h}^{2}V_{2},
\end{eqnarray}
where $A$ denotes the area of the horizon and $V_{2}=\int dxdy$.

Put the ansatz \eqref{5} and \eqref{6} into equations \eqref{3}, the independent equations of motion read
\begin{eqnarray}\label{9}
\rho''+(\frac{a'}{2}+\frac{f'}{f}) \rho'-\frac{m^2+4 J p^2 e^{-a}}{r^2 f} \rho-\frac{B}{r^2 f}&=&0,\nonumber\\
(m^2-\frac{4 J \rho^2}{r^4}) p-\phi'&=&0,\nonumber\\
\phi''+(\frac{2}{r}-\frac{a'}{2})\phi'+\lambda^2(\frac{p a'}{8}-\frac{p'}{4}-\frac{p}{2 r})&=&0,\nonumber\\
a'-\frac{\lambda^2 \rho'^2}{2 r^3}&=&0,\nonumber\\
f'+\frac{e^{-a}}{r}(\frac{4 J \rho^2}{r^4}-m^2)(\frac{4 J \rho^2}{r^4}-m^2+\frac{\lambda^2}{4})p^2\\
-\frac{3}{r}+(\frac{3}{r}+\frac{\lambda^2 \rho'^2}{4 r^3}) f+\frac{\lambda^2 \rho(m^2 \rho+2 B)+4 B^2}{4 r^5}&=&0,\nonumber
\end{eqnarray}
where a prime denotes the derivative with respect to $r$.
The first three equations are for the polarization field and Maxwell field, and
the last two are the two independent components of gravitational field equations.
In fact, there are three nonzero components of gravitational field equations,
but only two of them are independent due to the Bianchi identity.



We are interested in the black brane configurations which have a regular event horizon located at $r_{h}$.
Therefore, in addition to $f(r_{h})=0$, one must require $\phi(r_{h})=0$ in order for $g^{\mu\nu}A_{\mu}A_{\nu}$ being finite at the horizon.
We require the regularity conditions at the horizon $r=r_{h}$,
which means that all functions will have finite values at $r_{h}$ and admit
a series expansion in terms of $(r-r_{h})$.
Then, at the horizon, we have the following relations,
\begin{eqnarray}\label{12}
\phi&=&0,~\phi'=(m^2-\frac{4 J \rho^2}{r^4}) p,\nonumber\\
\rho'&=&\frac{\rho(m^2+4 J p^2 e^{-a})+B}{f'},\\
f'&=&-e^{-a}(4 J \rho^{2}-m^2)(4 J \rho^2-m^2+\frac{\lambda^2}{4}) p^2\nonumber\\
&~&+3-\frac{\lambda^2 \rho(m^2 \rho+2 B)}{4}+B^2.\nonumber
\end{eqnarray}
The black brane solution should be  asymptotically AdS.  Thus one has the following asymptotic solutions near the AdS boundary,
\begin{eqnarray}\label{13}
\rho&=&\rho_{+}r^{(1+\delta)/2}+\rho_{-}r^{(1-\delta)/2}+\cdots-\frac{B}{m^2},~a=a_{0}+\cdots~,\nonumber\\
\phi&=&\mu-\frac{\sigma}{r}+\cdots,~~f=1+\frac{f_{0}}{r^3}+\cdots,~p=\frac{\sigma}{m^2 r^2}+\cdots.\nonumber\\
\end{eqnarray}
where $\delta=\sqrt{1+4m^2}$ and the dots stand for the higher order terms of $1/r$,
$\rho_{\pm}$ and $f_{0}$ are all constants. The Breitenlohner-Freedman (BF) bound requires $m^2>-\frac{1}{4}$ according to the asymptotic solution of $\rho$.
According to the AdS/CFT dictionary, up to a normalization,
the coefficients $\mu$ and $\sigma$ are directly related to the chemical potential and charge density in the dual system, respectively. 
The magnetic moment density in the dual theory is defined  by the integration~\cite{29,Cai:2015bsa}
\begin{equation}\label{defN1}
N=-\frac{\lambda^2}{2}\int_{r_h}^\infty\frac{\rho e^{a/2}}{r^2}dr.
\end{equation}

Note that there is an additional restriction on $m^2$ when we treat the magnetic field B as the source of $\rho$.
In order to keep $\rho$ damping, we have to impose boundary condition for $\rho$ such that $\rho_{+}=0$
and a restriction on the parameter such that $m^2>0$~\cite{Cai:2015bsa}, otherwise the integration \eqref{defN1} will diverge or the equation for $\rho$ can not be linearized near the AdS boundary.
These two restrictions can be understood in another way:
when $B=0$, the constant $\rho_{+}$ should be viewed as the source of the corresponding
operator in the boundary field theory,
according to AdS/CFT duality. In order the symmetry to be broken spontaneously, one has to turn off the source term.

In addition, in order to make  $\rho$ condense in the case without
external magnetic field when the temperature is low enough,
the model parameters should violate the BF bound for $\rho$ near the horizon,
whose  geometry is an  AdS$_{2}$ for an extremal black brane.

To see how this requirement restricts the parameters, we first consider the solution for \eqref{9} in the case of $\rho(r)=B=0$ and $a(r)=V(M)=0$. The equations ($9$) become,
\begin{eqnarray}\label{15}
m^2p-\phi'&=&0,\nonumber\\
\phi''+ \frac{2}{r} \phi'-\lambda^2 (\frac{p'}{4}+ \frac{p}{2r})&=&0,\\
f'+\frac{3}{r}+\frac{p^2 m^2}{r}(m^2-\frac{\lambda^2}{4})-\frac{3}{r}&=&0.\nonumber
\end{eqnarray}
The equations admit an analytical solution  when $m^2\neq0$, i.e.,
\begin{eqnarray}\label{16}
\phi(r)&=&\mu(1-r_h/r),~~~~~~~p(r)=\frac{\mu r_h}{m^2 r^2},\\
f(r)&=&1-\frac{(\tilde{\mu}+1)r_h^3}{r^3}+\frac{\tilde{\mu}r_h^4}{r^4},~~~~ \tilde{\mu}=\frac{\mu^2}{r_h^2}(1-\frac{\lambda^2}{4 m^2}).\nonumber
\end{eqnarray}
Then the temperature of the black brane is
\begin{eqnarray}\label{17}
T&=&\frac{r_h}{4 \pi}(3-\tilde{\mu}).
\end{eqnarray}
Compared with the usual planar AdS RN black brane solution~\cite{Cai:1996eg}, we see that  $\mu^2$ in the metric of usual AdS RN solution is just replaced by $\tilde {\mu}$ in the new solution.
Namely, the new solution has the same form as the  planar AdS RN solution with the same Maxwell field $\phi(r)$,  but also with a nontrivial profile of $p(r)$ for the massive 2-form field.

\begin{figure}
\begin{center}
\includegraphics[width=0.15\textwidth]{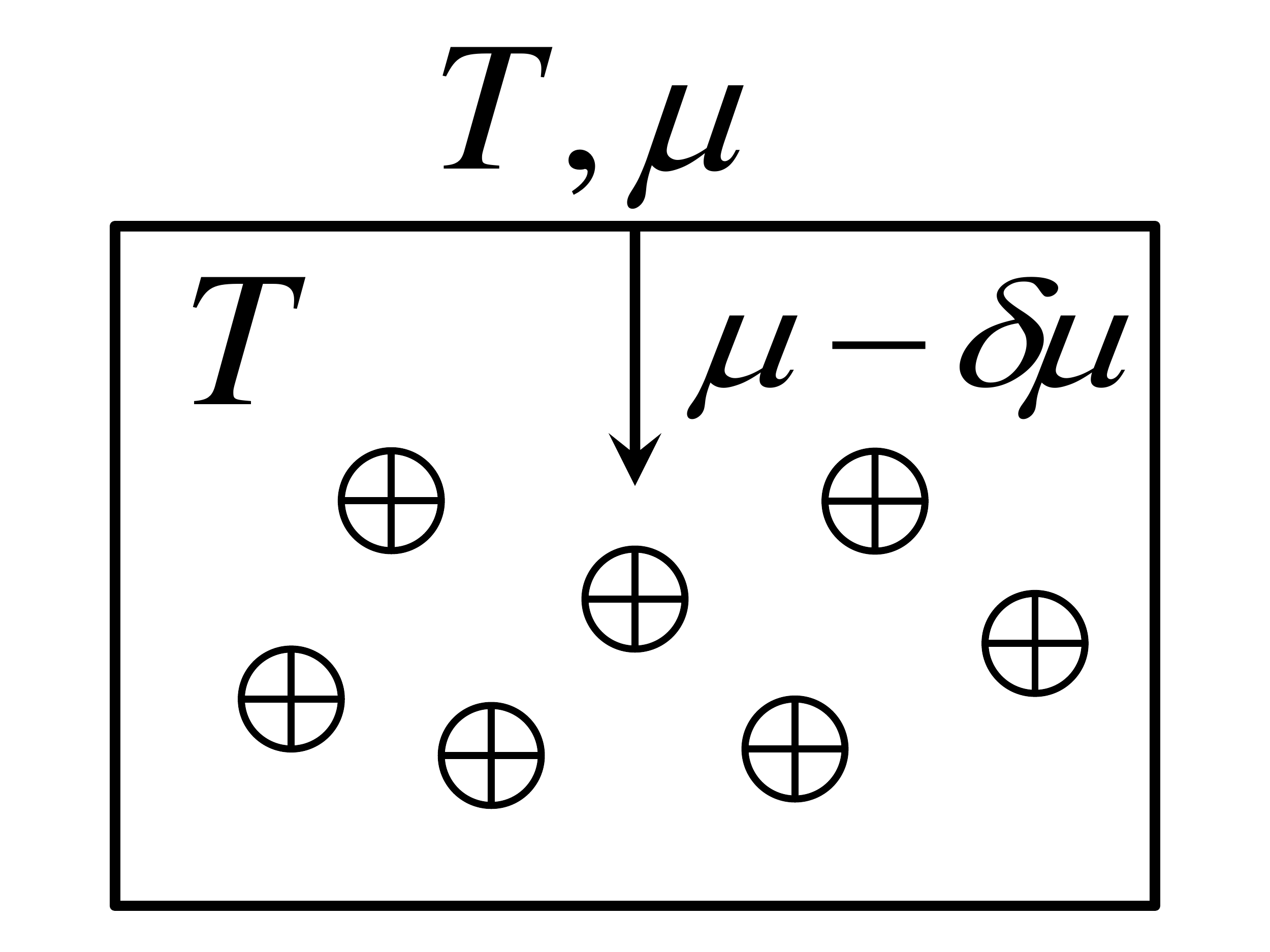}
\caption{The schematic diagram for chemical instability. }
\label{ins1}
\end{center}
\end{figure}

Then we can easily calculate the free energy $\Omega$ and the charge density $\sigma$ of the system as~\footnote{The total charge density $\sigma$ can be also calculated from the second equation in (\ref{3}), which gives the same value as the one in (\ref{freeE1}). This also confirms that the chemical potential of the system is  given by $\mu$.},
\begin{equation}\label{freeE1}
\begin{split}
\Omega&=\Omega(T,\mu)=-2V_2r_h^3(1+\widetilde{\mu})\\
\sigma&=\sigma(T,\mu)=-\frac2{V_2}\left(\frac{\partial\Omega}{\partial\mu}\right)_T =2(1-\frac{\lambda^2}{4 m^2})r_h\mu.
\end{split}
\end{equation}
We see that the properties of this black hole solution depends on the value of $1-\lambda^2/4 m^2$.
When $1-\lambda^2/4 m^2=0$, the solution is just  the planar AdS Schwarzschild black brane with nonzero U(1) gauge field and polarization field. To investigate the physical properties when $1-\lambda^2/4 m^2\neq0$, it is useful to compute the partial derivative of charge density with respect to chemical potential,
\begin{equation}\label{dsigmamu}
\left(\frac{\partial\sigma}{\partial\mu}\right)_T=-\frac1{V_2}\left(\frac{\partial^2\Omega}{\partial^2\mu}\right)_T=6 r_h(1-\frac{\lambda^2}{4 m^2})\frac{1+\widetilde{\mu}}{3+\widetilde{\mu}}.
\end{equation}
It is easy to see that Eq.~\eqref{dsigmamu} is positive when $1-\lambda^2/4 m^2>0$. However, when $1-\lambda^2/4 m^2<0$, the value of \eqref{dsigmamu} can become negative, which leads the dual boundary system to be chemical instability. To understand it, one can image that a box is submerged into the environment with fixed temperature $T$ and chemical potential $\mu$ in equilibrium state. The energy and charge can exchange between the interior of the box and environment through the wall (see Fig.~\ref{ins1}). Now because of thermal fluctuation, the chemical potential in the box then is $\mu-\delta\mu<\mu$, which leads to some charge coming into the box. However, because  Eqs.~\eqref{dsigmamu} is negative, adding the charge of the system will decrease the chemical potential of the system.
Then we see that the chemical potential will decrease again, which will lead that more charge come into the box.
So  the charge inside the box will be more and more and the system is unstable. From Eq.~\eqref{dsigmamu},
we can find the dual boundary system is chemical stable region only when temperature and chemical potential
satisfy\footnote{It is worth noting that this condition is only valid  in grand canonical ensemble. In other ensembles, the stable condition is different in general.},
\begin{equation}\label{cheuns1}
\frac{\sqrt{2}\pi}5<\frac{|\mu|}{(\frac{\lambda^2}{4m^2}-1)T}<\pi.
\end{equation}
If Eq.~\eqref{dsigmamu} is positive, we see that adding charge will increase the chemical potential, so after some time, the system can be in equilibrium again. In fact, in the case of $1-\lambda^2/4 m^2>0$, the solution shares all the properties of usual AdS RN  black brane except we have to replace  $\mu^2$ by  $\widetilde{\mu}$. But in the case of $1-\lambda^2/4 m^2<0$, the properties are very different from those of usual  AdS RN black brane. For example, the temperature gets increased when we increase the chemical potential while fixing the horizon radius and there does not exist zero temperature entropy.

In this paper, we only focus on the situation of $1-\lambda^2/4 m^2>0$, which can give a chemical stable dual boundary system and zero temperature black hole solution whose IR geometry has the form of asymptotic AdS$_{2}$ geometry. With  this, we consider the possibility of spontaneous symmetry breaking of this system in low temperature. This can be analyzed by solving the following equation in the background \eqref{16},
\begin{eqnarray}\label{18}
\rho''+\frac{f' \rho'}{f}-(\frac{m^2}{r^2 f}+\frac{4 J \mu^2}{r^6 m^4 f})\rho&=&0.
\end{eqnarray}
The condition of instability for $\rho$ in some low temperature is that the BF bound in AdS$_{4}$ is retained
but the BF bound in  AdS$_{2}$ is violated.
Near the horizon for an extremal black brane, the geometry is asymptotic AdS$_{2}$. In this region,
the asymptotic solution for equation \eqref{15} is,
\begin{eqnarray}\label{19}
\rho&=&\rho_{+}r^{-(1-\sqrt{1+4 \tilde{m}^{2}})/2}+\rho_{-}r^{-(1+\sqrt{1+4 \tilde{m}^2})/2},\nonumber\\
\tilde{m}^2&=&\frac{m^2}{6}+\frac{8 J}{m^2(4 m^2-\lambda^2)}.
\end{eqnarray}
So the conditions for spontaneous condensate  are,
\begin{eqnarray}\label{20}
&&m^2>0,~~~~~m^2(4 m^2-\lambda^2)>0,\\
&&\tilde{m}^2=\frac{m^2}{6}+\frac{8 J}{m^2 (4 m^2-\lambda^2)}<-\frac{1}{4}.\nonumber
\end{eqnarray}
One can immediately see that these inequalities  have solution only when $J<0$.

Under the restriction of \eqref{20}, there is a critical temperature, lower than which the nonzero $\rho$ begins to appear. Then the full coupled equations of motion do not admit an analytical solution. Therefore, we have to solve them numerically.
We will use shooting method to solve equations \eqref{9}.
In order to find the solutions for all the five functions $\mathcal{F}=\{\rho, \phi, p, a, f\}$, where we can
 directly solve $\phi(r)=\int_{r_h}^{\infty} dr(m^2-4 J\rho^2/r^4)p$ and then put it into the equation of $\phi''$ and obtain an equation for $p$,
we must impose suitable boundary conditions at both the horizon $r=r_{h}$ and conformal boundary $r\rightarrow\infty$.
There are two kinds of scaling symmetries which are useful when we perform numerical computations:
\begin{eqnarray}\label{10}
r\rightarrow\alpha r,~(t,x,y)\rightarrow(t,x,y)/\alpha,\\
(\rho,B)\rightarrow\alpha^{2}(\rho,B),~\phi\rightarrow\alpha\phi,~T\rightarrow \alpha T.
\end{eqnarray}
and
\begin{eqnarray}\label{11}
a\rightarrow a+\lambda,~e^{a}\rightarrow e^{\lambda} e^{a},~\rightarrow e^{-\frac{\lambda}{2}}t,~\phi\rightarrow e^{\frac{\lambda}{2}}\phi.
\end{eqnarray}
Under the above two scaling symmetries, we finally have four independent parameters $\{a(r_h),\rho(r_{h}),p(r_{h})\}$ and horizon radius $r_h$ at hand. In this paper, we will fix the boundary chemical potential to be unitary. When these four parameters are given, we can integrate the equations out of the horizon to obtain the whole solutions. When we perform the numerical computations, we can first set $\{r_h=1,a(r_h)=0\}$. Then for a given $\rho(r_h)$, we can adjust the value of $p(r_h)$ to match the boundary condition $\rho_+=0$ at $r\rightarrow\infty$. After solving the coupled differential equations, one should use the second scaling symmetry \eqref{11} to satisfy the asymptotic conditions $a(\infty)=0$. Then we  use the first transformation \eqref{10} to fix the chemical potential for each solution the same. By this method, we get one-parameter solution $\rho(r;T)$ for $T<T_c$. And then we can get the behavior of magnetic moment density $N$ with respect to  temperature $T$.

When the parameters satisfy the inequalities  \eqref{20},
the solution of $\rho\neq0$ will appear even in the case of $B=0$.
Because under the time reversal transformation,
$B\rightarrow-B$,  in order to make action be invariant, we have to have  $\rho\rightarrow-\rho$.
So when $B=0$ but $\rho\neq0$ in the source free case,
the time reversal symmetry is broken spontaneously.

In order to see the main properties of the model,
let us first study the model in probe limit for  simplicity. Here we may consider two kinds of probe limit.
The first one is to take the model parameter $\lambda\rightarrow 0$ as we did in Ref.~\cite{29}.  This case is just to neglect the back reaction of the massive 2-form field to the black brane
geometry and the Maxwell field, but considers the effect of the Maxwell field to the background geometry. This probe limit corresponds to the case in such materials that their electromagnetic response properties are very weak compared with the external field and have little effects on the their transport properties. In other words, in this probe limit, the influence of external field on the materials is considered, but we neglect the back reaction of electromagnetic response to the external field and structures of materials such as crystal structure or energy band.
The other  is to neglect all back reaction of matter fields including the Maxwell field to the background geometry. In this probe limit, the interaction between the electromagnetic response and external field is taken into account so that we can study how spontaneous magnetization influences the electric transport, but they both have little influence on the structures of materials.
We will study  two cases in the following sections separately.

\section{Probe limit in the case of $\lambda\rightarrow0$}
\label{prob1}
\subsection{Spontaneous magnetization and susceptibility}
Let us first investigate the spontaneous magnetization in the limit of $\lambda\rightarrow0$.
In this limit, we  neglect the back reaction of polarization field to the gauge field and background geometry.
The background geometry and  the Maxwell field can be taken as
\begin{eqnarray}\label{21}
\phi(r)&=&\mu(1-1/r),~~~f(r)=1-\frac{1+\mu^2}{r^3}+\frac{\mu^2}{r^4},\nonumber\\
T&=&\frac{1}{4 \pi}(3-\mu^2),
\end{eqnarray}
where we have set  the horizon radius $r_h=1$.
In this case, the equations of the polarization field read
\begin{eqnarray}\label{22}
\rho''+\frac{f'}{f}-\frac{m^2+4 J p^2}{r^2 f}\rho&=&\frac{B}{r^2 f},\nonumber \\
(m^2-\frac{4 J \rho^2}{r^4})p-\frac{\mu}{r^2}&=&0.
\end{eqnarray}
Note that the expression of $p(r)$ can be solved directly. We  put it into the equation of $\rho(r)$, and get
\begin{equation}\label{23}
\rho''+\frac{f'}{f}\rho'-[\frac{m^2}{r^2 f}+\frac{4 J \mu^2}{(m^2r^2-\frac{4 J \rho^2}{r^2})^2r^2f}]\rho-\frac{B}{r^2 f}=0.
\end{equation}
This equation shows that $\rho$ will be spontaneously condensed below a critical temperature only when $m^2>0$.
When $m^2>0$, near the critical temperature where $\rho$ is very small and therefore  the term $\rho^2$ can be neglected in (\ref{23}), thus increasing the chemical potential\footnote{This implies  that  the temperature is decreased in grand canonical ensemble [see (\ref{21})].} will decrease the effective mass at the horizon (note that $J<0$), which leads that $\rho$ can be condensed spontaneously below a critical temperature.~\footnote{If $m^2<0$,  $\rho$ will get spontaneously condensed at a temperature higher than a critical temperature. This solution is unstable and we will not
discuss it any more.} 
Furthermore, the restrictions on the parameters in the case of $\lambda\rightarrow0$ are,
\begin{eqnarray}\label{24}
\tilde{m}^2&=&\frac{m^2}{6}+\frac{2 J}{m^4}<-\frac{1}{4},~~~~~m^2>0.
\end{eqnarray}
In what follows, since different parameters will give similar results,
we will choose $m^2=-J=1/8$ as a typical example when we perform numerical computations.

To study the spontaneous magnetization, we need set $B=0$. Since near the critical temperature,  $\rho$ is very small, the nonlinear term of $\rho$ can be neglected in (\ref{23}).To find the critical temperature, we can solve the linearized equation of $\rho$  with initial condition
\begin{eqnarray}\label{25}
\rho'&=&\frac{m^6+4 J(3-4 \pi T)}{4\pi T m^4}\rho,
\end{eqnarray}
at the horizon $r=1$ and boundary condition $\rho_{+}=0$ at $r\rightarrow\infty$.  Without loss the generality, we take the initial value of $\rho$ at horizon to be unity,
and treat $T$ as the shooting parameter to match the source free condition.
There will be many solutions for shooting parameters $T$,
 we choose the highest one as the critical temperature $T_{c}$.
We find the critical temperature $T_{c}/\mu\simeq1.78$ for the case of given model parameters.

\begin{figure}
\begin{center}
\includegraphics[width=0.3\textwidth]{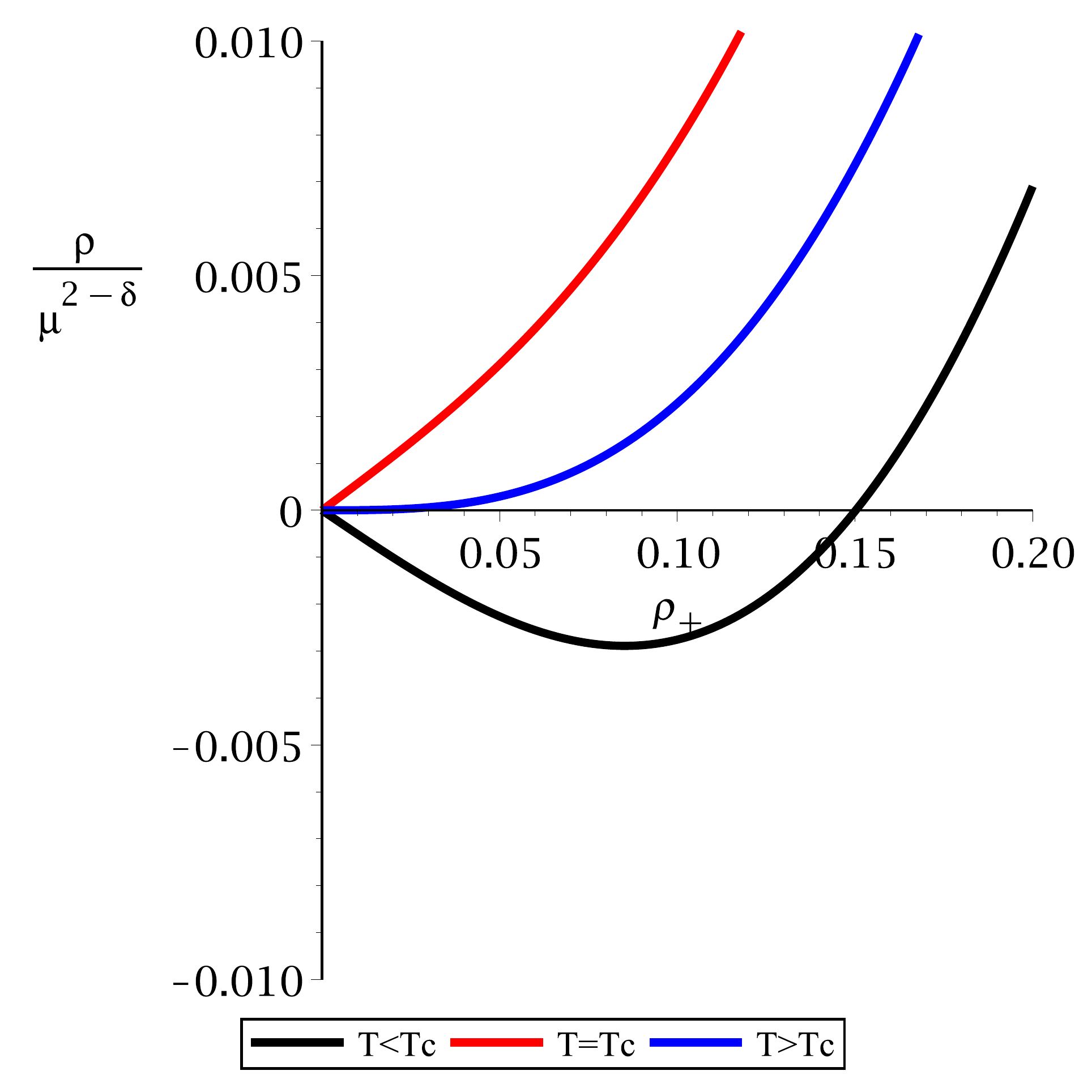}
\includegraphics[width=0.25\textwidth]{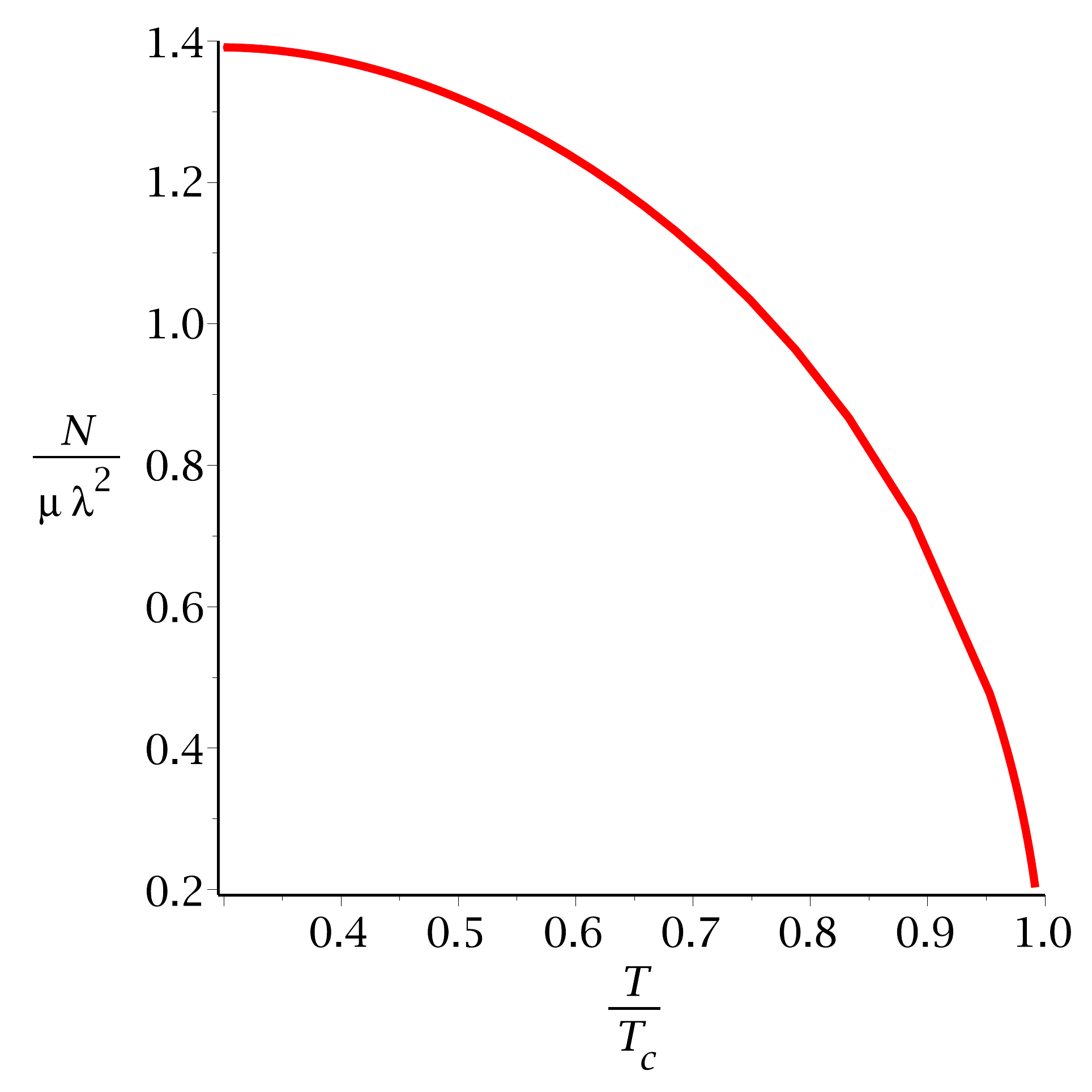}
\caption{Top panel: The value of $\rho(r_h)$ in  different temperature. Bottom panel: The magnetic moment $N$ as a function of  temperature.
Here we choose the model parameters as $m^2=-J=1/8$. The critical temperature $T_{c}/\mu\simeq1.78$. } \label{F1}
\end{center}
\end{figure}

When the temperature is lower than  the critical temperature $T_{c}$,
in order to examine whether the polarization field $\rho$
can make spontaneous magnetization
when the external magnetic field $B=0$.
We plot the value of $\rho_{+}$ versus $\rho(r_{h})$ at the horizon in the top panel of Fig.~\ref{F1}.
Each curve in this plot corresponds to different temperature.
This plot  shows a typical example in the high and low temperature cases
which correspond to the red and the black lines, respectively, while the case with  the critical temperature is
shown by the blue curve.
In the case of high temperature, we see that the curve has no intersecting point with horizontal
axis except a trivial point at the origin, which corresponds to a trivial solution with $\rho=0$ and it describes the paramagnetic phase.
On the other hand, when the temperature is low enough,
we find that there exists a nontrivial solution which locates at $\rho(r_{h})\neq0$.
This solution breaks the time reversal symmetry and the system gets into a ferromagnetic phase.
As a result,  we see that the model indeed can give rise to a paramagnetism/ferromagnetism phase transition
in the case without external magnetic field.


Furthermore, when the temperature is lower than the critical temperature $T_{c}$,
we have to solve the equation \eqref{23} with the condition of $\rho_{+}=0$ by shooting method.
The spontaneous magnetization $N$ is defined by the integration \eqref{defN1} with $a=0$.
The spontaneous magnetic moment with respect to temperature is shown in the bottom panel of Fig.~\ref{F1}.
In addition, since the magnetic moment of the polarization field obtains an expectation value, the time reversal symmetry is broken spontaneously.

By fitting this curve near the critical temperature,
we find that there is a square root behavior for the magnetic moment versus temperature,
which is a typical behavior for a second order phase transition,
specifically, for $m^2=-J=1/8$, we have
\begin{eqnarray}\label{27}
N^{2}/\lambda^{4} \mu^{2}\simeq4.910(1-T/T_{c}).
\end{eqnarray}
This gives the critical exponent $1/2$, the same as the one from the mean field theory.

Except for the magnetic moments which is one of the characteristic properties of ferromagnetic material,
another remarkable one is the behavior of susceptibility density of the material in the external magnetic field.
The static susceptibility density is defined by
\begin{eqnarray}\label{28}
\chi&=&\lim_{B\rightarrow0} \frac{\partial N}{\partial B}.
\end{eqnarray}
When we turn on the external magnetic field $B$, the function $\rho$ is nonzero in any temperature.
In order to compute the susceptibility density, we need to
shoot for the boundary conditions with one parameter $\rho(r_h)$ for equation \eqref{23} under the given external magnetic $B$ and temperature $T$.
From the definition of $\chi$, which involves only the behavior of $B\rightarrow 0$,
so the computation can be simplified in the following way.


\begin{figure}
\centering
\includegraphics[width=0.35\textwidth]{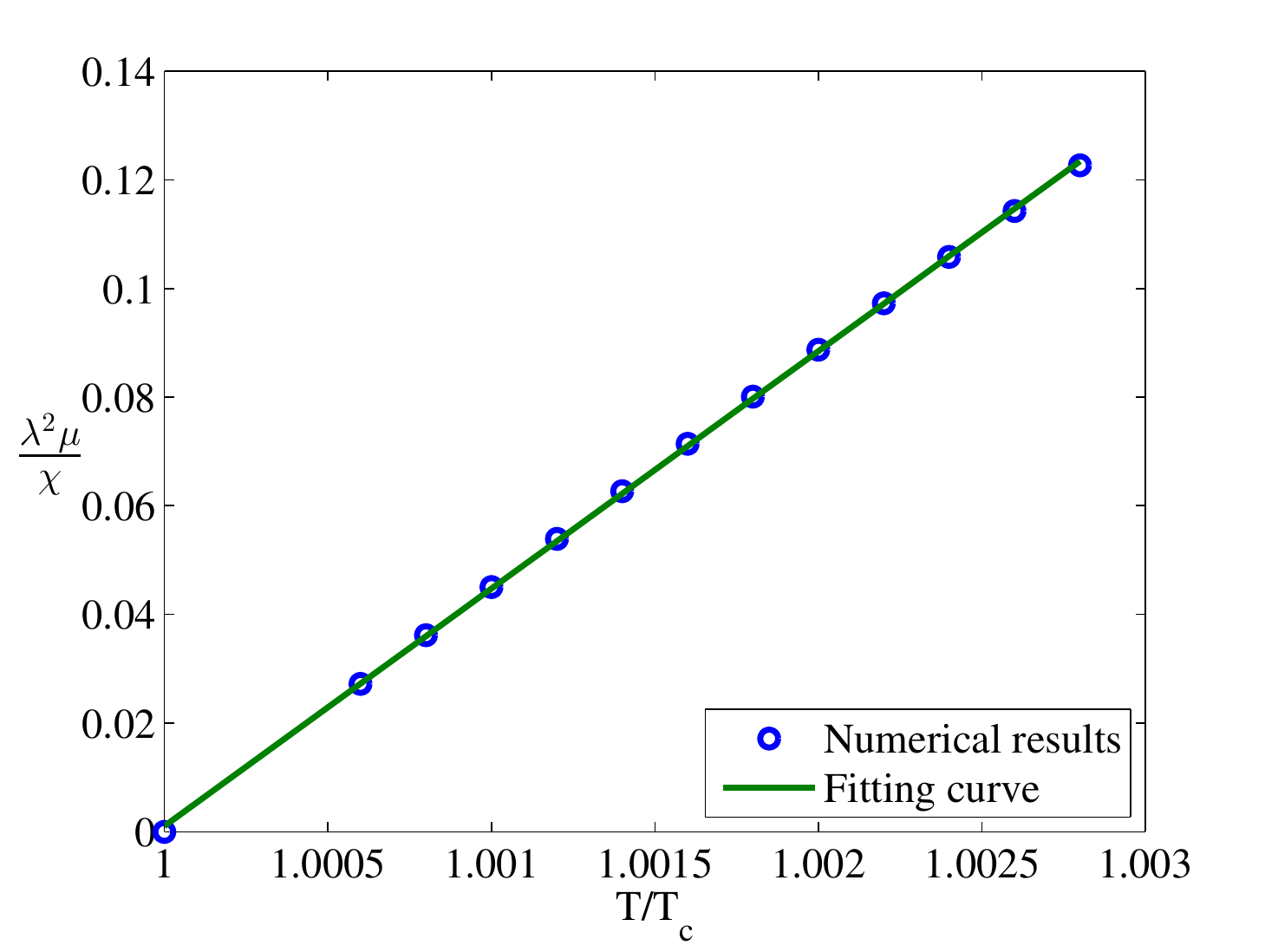}
\caption{The behavior of the inverse susceptibility density in the paramagnetic phase near the critical temperature when $m^2=-J=1/8$. Here we set $2\kappa^2=1$ for convenience.} \label{F2}
\end{figure}

In the case of $T>T_{c}$, considering the result $\rho=0$ when $B=0$,
we except that $\rho$ is in the same order as $B$,
then  $\rho^{2}$ is as the same order as $B^{2}$.
So in the case with weak magnetic field, we can neglect nonlinear terms in equation \eqref{23}. In that case, we just need to solve the ``linearized"  equation
\begin{eqnarray}\label{29}
\rho''+\frac{f' \rho'}{f}-(\frac{m^2}{r^2 f}+\frac{4 J \mu^2}{r^6 m^4 f})\rho-\frac{B}{r^2 f}&=&0.
\end{eqnarray}
Because $\rho\propto B$, we can set $B=1$, Solving the equation \eqref{29} with the boundary condition $\rho_{+}=0$,
the susceptibility density then can be computed by
\begin{eqnarray}\label{30}
\chi&=&N=-\lambda^2 \int^{\infty}_{r_h} \frac{\rho}{2 r^{2}} dr.
\end{eqnarray}
In Fig. \ref{F2}, we show the magnetic susceptibility as a function of temperature and
find it satisfies the Curie-Weiss law of ferromagnetism in the region of $T\rightarrow T_{c}^+$.
Concretely, for the chosen model parameters, we have
\begin{eqnarray}\label{31}
\lambda^{2}/\mu \chi&\simeq&4.0499(T/T_{c}-1).
\end{eqnarray}
So we can conclude that the dual system is in a paramagnetic phase in high temperature and
ferromagnetic phase in low temperature. The model can describe a paramagnetism/ferromagnetism phase transition. We will see later that the exponents in Eqs.\eqref{27} and \eqref{31} are exact and can be obtained by analytical methods.

\subsection{Holographic Ginzburg-Landau formulation}
In  action \eqref{1}, two parameters appear in the Lagrangian. One is the mass of the polarization field, which corresponds to the conformal dimension of the dual operator by the standard dictionary. The other  is the self-interaction coupling constant $J$. So a natural question is what the meaning of this constant  is in the dual boundary theory? In this subsection, we will try to give a quantitative interpretation. Since we are working in the probe limit, the geometry and external Maxwell field are fixed, this leads to a simplification to compute the partition function of the bulk theory. This is the reason that we consider this kind of probe limit here. Our method is to compute the effective grand thermodynamic potential in both sides and  to equate them (which are equivalent to compute the partition functions in both sides and to equate them). Then we can ``read off"  the meaning of $J$ in the dual boundary theory.

We assume that the the spontaneous magnetization in the boundary theory can be described by so-called Ising universality class model at least in the vicinity of critical temperature, whose Hamiltonian is,
\begin{equation}\label{Ising1}
H=-\frac12\sum_{r,r'}\mathcal{J}(r-r')\overrightarrow{s}(r)\cdot\overrightarrow{s}(r')+\sum_{r}K[\overrightarrow{s}(r)^2-1]^2,
\end{equation}
where $r, r'$ are the positions of lattices, $\mathcal{J}(r-r')>0$ is the exchange integration of lattices $r$ and $r'$, $s(r)$ the z-component of spin at lattice  $r$, $K$ is a positive constant which corresponds to the deviation from the Ising model. When $K\rightarrow\infty$, Hamiltonian \eqref{Ising1} reduces to that of the usual Ising model.

In this paper, the dual boundary is  a (2+1)-dimensional spacetime, i.e.,  a film system. Then the spin (or magnetic moment) in fact is just a pseudoscalar. So we have $\overrightarrow{s}=s$. To compare with our holographic model, it is convenient to change Hamiltonian \eqref{Ising1} into the one in continuous limit,
\begin{equation}\label{Ising2}
\begin{split}
H=&\int d^2xl^{-2}\left[-\frac12\mathcal{J}l^2R_{\mathcal{J}}^2s(x)\overrightarrow{\nabla}^2s(x)\right.\\
&\left.-(\mathcal{J}+2K)s(x)^2+Ks(x)^4\right],
\end{split}
\end{equation}
where $l$ is the lattice spacing and $\mathcal{J}=\sum_r\mathcal{J}(r)$, $R_{\mathcal{J}}^2=\sum_rr^2\mathcal{J}(r)$. The summations for $\mathcal{J}$ and $R_{\mathcal{J}}^2$ are only in one  crystal lattice. We see that the Ising like Hamiltonian is a $\lambda\phi^4$ theory if we use $\widetilde{s}=s\sqrt{\mathcal{J}}R_{\mathcal{J}}$. From  Appendix~\ref{app1}, we can find that the grand thermodynamic potential in the mean field approximation in the high temperature limit reads,
\begin{equation}\label{Veffs1}
\begin{split}
\widetilde{\Omega}\simeq&\frac{1}2(\frac{\lambda_s\gamma_ET}{4\pi}-m_s^2)\widetilde{s}_{cl}^2+\frac{\lambda_s\widetilde{s}^4_{cl}}{4!}+\cdots
\end{split}
\end{equation}
with
\begin{equation}\label{massS}
m_s^2=\frac{2\mathcal{J}+4K}{\mathcal{J}l^2R_{\mathcal{J}}^2},~~\lambda_s=\frac{4!K}{\mathcal{J}^2l^{2}R_{\mathcal{J}}^4}.
\end{equation}
Here we have used $\widetilde{s}_{cl}$ to represent the classical value of $s$. We see that there is a critical temperature $T_c=4\pi m_s^2/\lambda_s\gamma_E$. When $T>T_c$, the thermodynamic equilibrium phase corresponds to $\widetilde{s}_{cl}=0$, while  $T<T_c$, the thermodynamic equilibrium phase has $\widetilde{s}_{cl}\neq0$.

Now let us consider the gravity side of the holographic model to compute the grand thermodynamic potential. It is  convenient to make a coordinate transformation by $z=r_h/r$. As $p$ can be solved directly, then we  put it into the equation of $\rho(r)$, and get
\begin{equation}\label{Eoms1}
\begin{split}
\rho''+\left(\frac2z+\frac{f'}{f}\right)\rho'-[\frac{m^2}{z^2 f}+\frac{4 J \mu^2z^2}{(m^2-4 J\rho^2z^4)^2f}]\rho&=\frac{B}{z^2 f},\\
(m^2-4 J \rho^2z^4)p-\mu z^2&=0.
\end{split}
\end{equation}
As we will care about the behavior of $T\rightarrow T_c$, the value of $\rho$ will be a small quantity near the transition point. In this case, we can make a Taylor's expansion on the nonlinear term of $\rho$ in Eq.~\eqref{Eoms1} as,
\begin{equation}\label{Taylor1}
\frac{4 J \mu^2z^2}{(m^2-4 J\rho^2z^4)^2}=\frac{4 J \mu^2z^2}{m^4}+\frac{32J^2\mu^2\rho^2z^6}{m^6}+\mathcal{O}(\rho^4)
\end{equation}
Neglecting the high order terms,  Eq.~\eqref{Eoms1} can be rewritten as
\begin{equation}\label{rhop}
\begin{split}
&\widehat{L}\rho=\widetilde{J}_f\rho^3z^8-B,\\
&\widehat{L}=-\frac{d}{dz}\left[z^2f(z)\frac{d}{dz}\right]+q(z),\\
&q(z)=m^2+\frac{4 J \mu^2z^4}{m^4},\\
&\widetilde{J}_f=-32J^2\mu^2/m^6<0.
\end{split}
\end{equation}

Up to the order of $\rho^4$, the part of polarization field in action  \eqref{1} can be written as,
\begin{equation}
\begin{split}
\frac{S(T,B;\rho)}{\lambda^2V_2}&=\left.(\frac{z^2}2f\rho'\rho+zf\rho^2)\right|^{z_h}_{0}\\
&-\int_{0}^{z_h} dz\left[\frac {\rho}2\widehat{L}\rho+B\rho-\frac{\widetilde{J}_f}4z^8\rho^4\right],
\end{split}
\end{equation}
which is a function of $T$ and $B$, but a functional of $\rho$.
The asymptotic solution for~\eqref{rhop} is
\begin{equation}\label{symp1}
\rho=\widetilde{\rho}-\frac{B}{m^2},~\text{with}~\widetilde{\rho}=\rho_+z^{-(1+\delta)/2}+\rho_-z^{-(1-\delta)/2}.
\end{equation}
The source free condition is  $\rho_+=0$ as $z\rightarrow0^+$. Under this, the grand thermodynamic potential or free energy in grand canonical ensemble $\Omega$ is,
\begin{equation}\label{action1}
\begin{split}
\Omega(T,B;\rho)&=\widetilde{\Omega}(T,B;\rho)V_2\\
&=\lambda^2V_2\int_{0}^{z_h} dz\left[\frac {\rho}2\widehat{L}\rho+B\rho-\frac{\widetilde{J}_f}4z^8\rho^4\right].
\end{split}
\end{equation}
According to thermodynamic relationship,
\begin{equation}\label{thero1}
\begin{split}
&d\Omega(T,B)=-SdT-V_2NdB\\
&\Rightarrow N/\lambda^2=-\frac1V_2\left(\frac{\partial \Omega(T,B)}{\partial B}\right)_{T}.
\end{split}
\end{equation}
It seems that the magnetic moment should be,
\begin{equation}\label{defN0}
N=-\frac{\lambda^2}{V_2}\left(\frac{\partial \Omega(T,B;\rho)}{\partial B}\right)_{T,\rho}=-\lambda^2\int_{0}^{z_h} \rho dz.
\end{equation}
However, in our previous papers, we defined the magnetic moment as,
\begin{equation}\label{defN2}
N/\lambda^2=-\int_{0}^{z_h} \frac{\rho}{2} dz.
\end{equation}
There is a difference factor  $1/2$ between \eqref{defN2}  and \eqref{defN0}. Now we show that  the expression~\eqref{defN0} is not true.
The reason is as follows. The relation~\eqref{thero1} is on-shell, while the equation~\eqref{action1} is off-shell.
In order to obtain the differential relation of $\Omega(T,B;\rho)$ with respect to $(T,B,\rho)$, we need to use the Euler homogenous function theorem.
We should first note that under the scaling transformation $z\rightarrow kz,~(t,x,y)\rightarrow k(t,x,y)$, we have $\Omega(kT,k^2B;k^2\rho)=k\Omega(T,B;\rho)$,
which gives,
\begin{equation}\label{dfT}
\begin{split}
\Omega(T,B;\rho)=&\left(\frac{\partial \Omega}{\partial T}\right)_{\rho,B}T+2\left(\frac{\partial \Omega}{\partial B}\right)_{\rho,T}B\\
&+2\int_{0}^{z_h}\left(\frac{\delta \Omega}{\delta \rho(z)}\right)_{T,B}\rho(z)dz.
\end{split}
\end{equation}
Submitting \eqref{action1} into \eqref{dfT} and considering the on-shell condition $\delta \Omega/\delta\rho=0$, we find that,
\begin{equation}\label{defN}
N/\lambda^2=-\frac{1}{V_2}\left(\frac{\partial \Omega(T,B)}{\partial B}\right)_{T,\text{on-shell}}=-\int_{0}^{z_h} \frac{\rho}{2} dz.
\end{equation}
This is just  the definition  \eqref{defN2} .


The key step for computing the grand thermodynamic potential is to structure the Sturm-Liouville problem\footnote{The method is similar to the one used in Ref.~\cite{Yin:2013fwa}, but is completely different from the Sturm-Liouville (SL) eigenvalue method in Ref.~\cite{Siopsis:2010uq}, there the precision depends on the trial function one chooses.},
 which is the following ODE:
\begin{equation}\label{SL1}
\begin{split}
\widehat{P}\rho_n&=\frac{\widehat{L}\rho_n}{\omega(z)}\\
&=\frac1{\omega(z)}\left\{-\frac{d}{dz}\left[z^2f(z)\frac{d\rho_n}{dz}\right]+q(z)\right\}\rho_n=\lambda_n\rho_n.
\end{split}
\end{equation}
with the boundary conditions: \\
(a) At $z=z_h$, $f(z_h)=0$,  $|\rho_n(z_h)|$ is required to be finite;\\
(b) At $z\rightarrow0^+$, we impose $\rho_n(0)=0$.\\

The weight function $\omega(z)$ can be an arbitrary  positive continuous function in the region of $[0,z_h]$. From a practical point of view, we choose weight function such that the values of $\lambda_n$ will not influence the asymptotic behaviors of equation~\eqref{SL1} when $z\rightarrow0^+$. There are many choices for weight function. Here we choose $\omega(r)=z^k$ with an integer $k>2$.

Once note that the asymptotic solution when $r\rightarrow\infty$ for equation~\eqref{SL1} is,
\begin{equation}\label{asym1}
\rho_n=\rho_+z^{-(1+\delta)/2}+\rho_-z^{-(1-\delta)/2}
\end{equation}
one can find that the boundary condition (b) corresponds to $\rho_+=0$. Let $\mathcal{L}^2([0,z_h],\omega(z),dz)$ be the Hilbert space of square integrable functions on $[0,z_h]$, i.e.,
\begin{equation}\label{H1}
\begin{split}
&\mathcal{L}^2([0,z_h],\omega(z),dz)\\
=&\left\{h: [0, z_h]\mapsto\mathbb{R}\left| \int_{0}^{z_h}\omega(z)|h(z)|^2dz<\infty\right.\right\}
\end{split}
\end{equation}
with the inner product
\begin{equation}\label{inprod}
\langle h_1,h_2\rangle=\int_{0}^{z_h}\omega(z)h_1(z)h_2(z)dz,
\end{equation}
and $D$ be the subspace of $\mathcal{L}^2([0,z_h],\omega(z),dz)$ that satisfies the boundary conditions of (a) and (b), i.e.,
\begin{equation}\label{D1}
\begin{split}
D=&\left\{\forall h\in\mathcal{L}^2([0,z_h],\omega(z),dz)\left|h\in C^2[0,z_h]\right.,\right.\\
&\left.h(0)=0,~|h(z_h)|<\infty\right\}.
\end{split}
\end{equation}
Then we can prove that $\widehat{P}$ is the self-adjoint operator on $D$, i.e.,
\begin{equation}\label{slad}
\forall h_1,h_2\in D,\langle h_1,\widehat{P}h_2\rangle=\langle \widehat{P}h_1,h_2\rangle.
\end{equation}
According to the properties of SL problem, the solutions of~\eqref{SL1} form a function basis on $D$  with which one can expand any functions belonging to $D$, i.e.,
\begin{equation}\label{expand0}
\langle\rho_n,\rho_k\rangle=\delta_{nk},
\end{equation}
and
\begin{equation}\label{expand0b}
\forall h\in D, \exists\{c_n\}\subset\mathbb{R}, h(z)=\sum_{n=1}^\infty c_n\rho_n(z)
\end{equation}
with $c_n=\langle\rho_n,h\rangle$.

\begin{figure}
\includegraphics[width=0.23\textwidth]{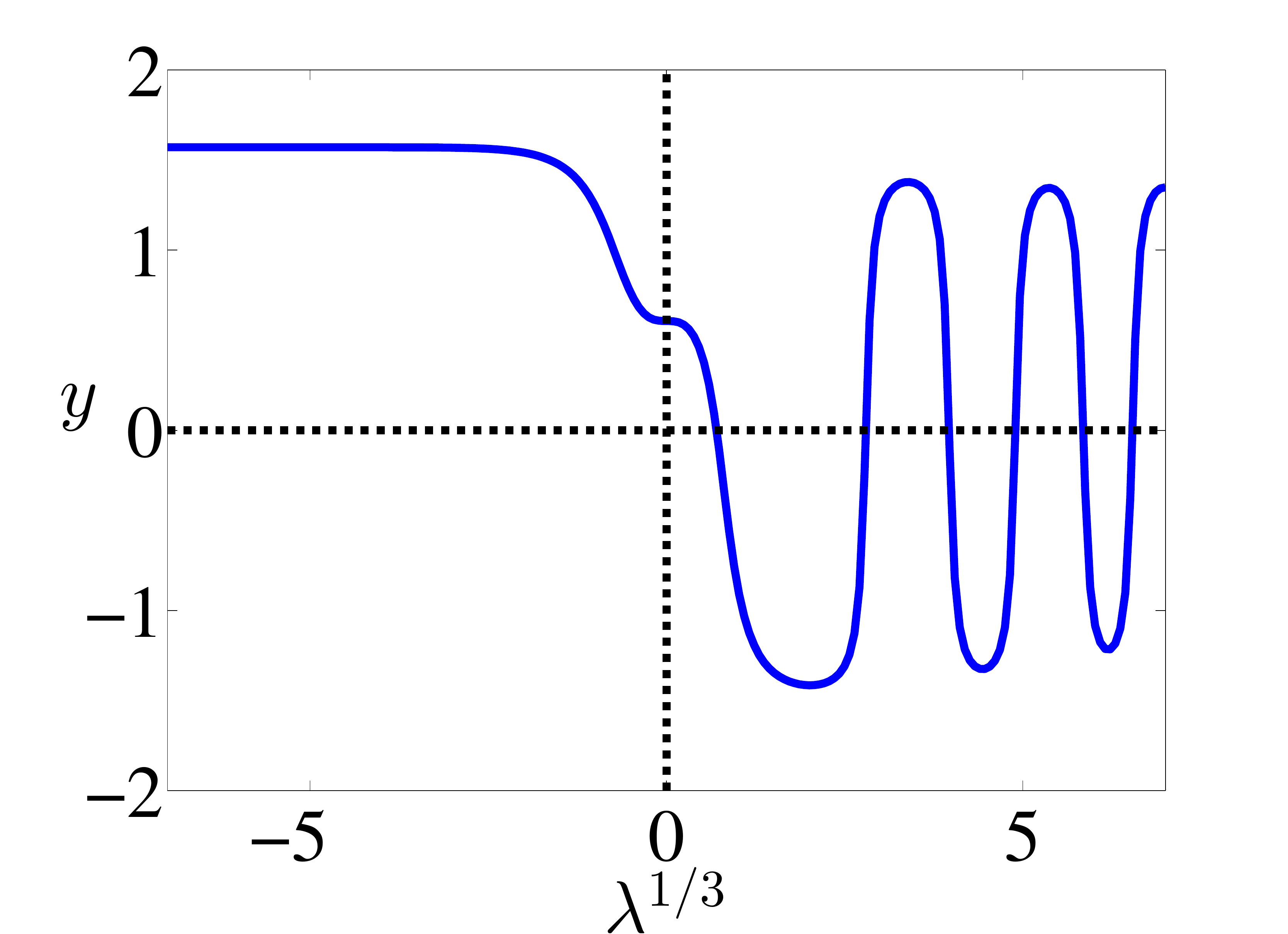}
\includegraphics[width=0.23\textwidth]{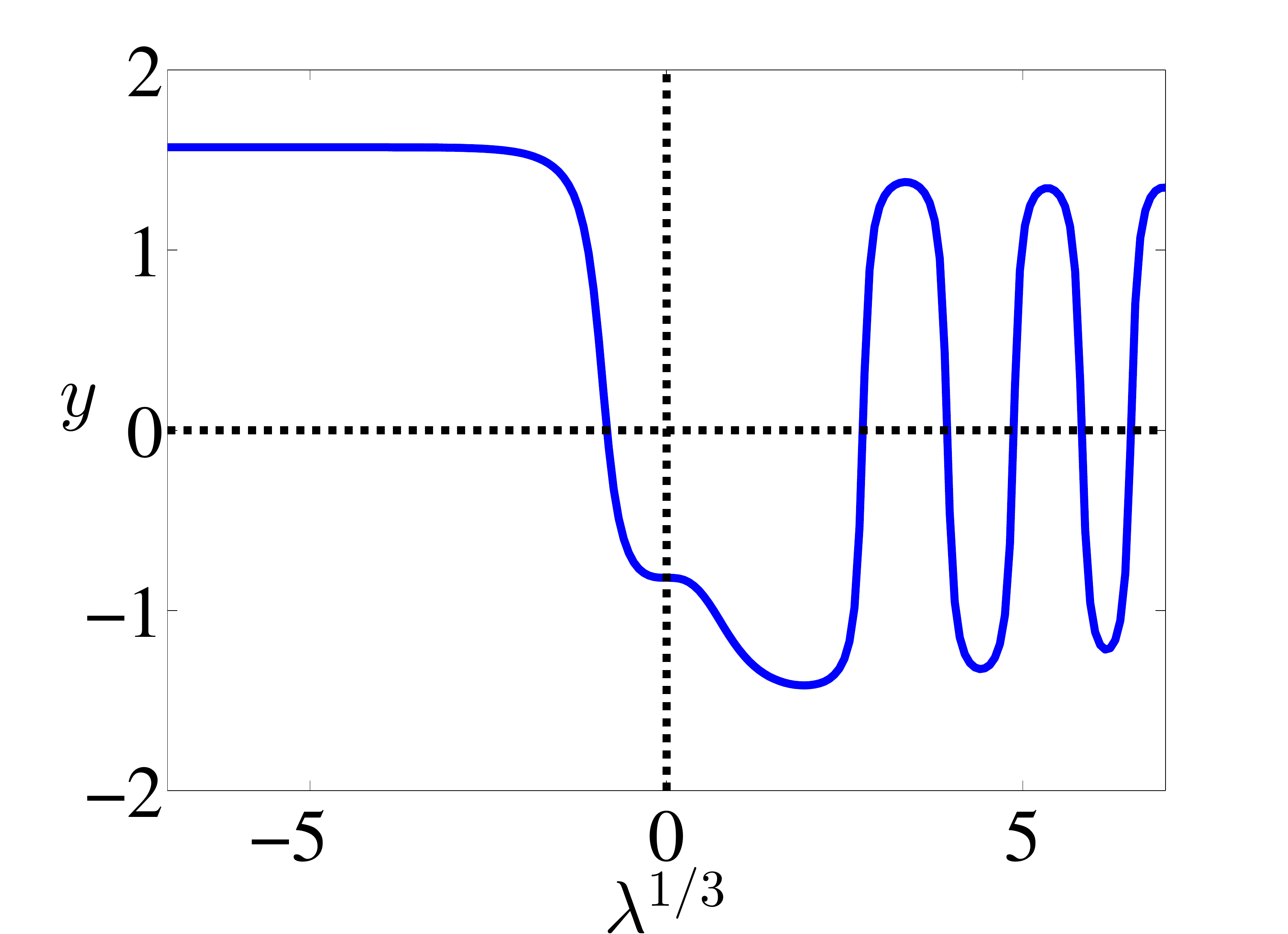}
\caption{The function $y(\lambda)=\arctan[10\rho_+(\lambda)]$ with respect to $\lambda$ at temperature of $T\simeq2.00T_c$(left one) and $T\simeq0.491T_c$(right one) in the case of $m^2=1/8$. The x-coordinate values of zero points give the eigenvalues $\lambda_n$.  Here $T_c\simeq1.7766\mu$.}
\label{lambdarho}
\end{figure}

Note that all the parameters in equation~\eqref{SL1} depend on temperature, so do the eigenvalues $\lambda_n$ and eigenfunctions $\rho_n$. Indeed, the minimal eigenvalue, i.e., the first eigenvalue $\lambda_1$ is the function of temperature. There is a critical temperature $T_c$, at which we have $\lambda_1=0$. We can find that when $T>T_c$, $\lambda_1>0$. In order to show this,
we take $m^2=1/8$ as an example to plot  $\rho_+$ with respect to $\lambda$.
The  boundary condition (b) is equivalent to $\rho_+(\lambda_n)=0$. Fig.~\ref{lambdarho} shows the value of $y(\lambda)=\arctan[10\rho_+(\lambda)]$ with respect
to $\lambda^{1/3}$ (it is just for convenience). The $x$-coordinate values of zero points ($y=0$) give the eigenvalue $\lambda_n$.
We can see that all the eigenvalues are positive when $T>T_c$ and $\lambda_1<0$ when $T<T_c$. For any temperature,
the equation has infinite eigenvalues and  the smallest eigenvalue exists, which shows the system is stable.
%

Let us now turn our attention to the free energy~\eqref{action1}. For convenience, we will use scaling transformation to set $z_h=1$ in the process of computation, and then transform into the case of fixing chemical potential in the final results.

Let $\widetilde{\rho}(r)=\rho(r)+B/m^2$ be any function configuration belonging to $D$, in which $\rho(r)$ dose't need to be the solution of EoM~\eqref{rhop}. We can use the eigenfunction $\rho_n$ to expand  $\widetilde{\rho}(r)$ and magnetic moment as,
\begin{equation}\label{exprho1}
\begin{split}
\widetilde{\rho}=\sum_{n=1}^\infty c_n\rho_n\Leftrightarrow\rho=\sum_{n=1}^\infty c_n\rho_n-\frac{B}{m^2},\\
 N=\frac{\lambda^2B}{2m^2}-\lambda^2\int_{0}^{1}\frac{\widetilde{\rho}}{2}dz=\frac{\lambda^2B}{2m^2}-\frac{\lambda^2}2\sum_{n=1}^\infty c_nN_n,
 \end{split}
\end{equation}
where $c_n$ and $N_n$ are coefficients,  defined as,
\begin{equation}\label{coeffN}
c_n=\int_{0}^{1}\omega\widetilde{\rho}\rho_ndz,~N_n=\int_{0}^{1}\rho_ndz.
\end{equation}
Here we have assumed that $\{\rho_n\}$ is an unit base. Then the variational principle of $\Omega(T,B;\rho)$ underlying the equations of motion, or finding a solution of EoM~\eqref{rhop}, is equivalent to  minimizing  $\Omega(T,B,c_n)$ with respect to $c_n$'s.

Let us  consider the case of spontaneous magnetization, i.e., the case with $B=0$. In this case, we have
\begin{equation}\label{action2}
\widetilde{\Omega}(T,c_n)=\lambda^2\int_{0}^{1} dz\left[\frac {\omega\rho}2\widehat{P}\rho-\widetilde{J}_fz^8\rho^4/4\right],
\end{equation}
with $\rho=\sum_{n=1}^\infty c_n\rho_n$.
Using the orthogonal relationship, we have,
\begin{equation}\label{expand2}
\begin{split}
\widetilde{\Omega}(T,c_n)&=\frac{\lambda^2}2\langle\rho, \widehat{P}\rho\rangle-\frac{\lambda^2\widetilde{J}_f}4\int_{0}^{1}z^8\rho^4dz\\
&=\frac{\lambda^2}2\sum_{n=1}^\infty \lambda_nc_n^2-\frac{\lambda^2\widetilde{J}_f}4\int_{0}^{1}z^8\rho^4dz.
\end{split}
\end{equation}
If $T>T_c$, then $\lambda_n>0$. Because of $\widetilde{J}_f<0$, we can find that $\widetilde{\Omega}(T, c_n)\geq0$. The minimization of $\Omega(T, c_n)=0$ is achieved only when $c_n=0$, i.e., $\rho=0$. So the nonzero solution appears only when $\lambda_1<0$, i.e., $T<T_c$. This is just what we have obtained in the pervious section.

When $T\rightarrow T_c^-$, we can set $\lambda_1=a_0(T/T_c-1)$ with $a_0>0$ and assume that the off-shell solution is dominated by the first term in \eqref{exprho1} only, i.e.,  $|c_1|\gg c_n$ for $n\geq2$ in \eqref{coeffN}. As  a result,  we have,
\begin{equation}\label{expand3}
\begin{split}
\lambda^{-2}\widetilde{\Omega}(T,c_n)&\simeq \frac12\lambda_1c_1^2-\frac{\widetilde{J}_fc_1^4}4\int_{0}^{1} dz\rho_1^4z^8,\\
&\simeq \frac12a_0(T/T_c-1)c_1^2-\widetilde{J}_fc_1^4 a_{1}
\end{split}
\end{equation}
with $a_1=\frac14\int_{0}^{1}\rho_1^4z^8 dz|_{T=T_c}>0$ and,
\begin{equation}\label{expandN2}
N\simeq -\lambda^2c_1N_1/2.
\end{equation}
Put \eqref{expandN2} into~\eqref{expand3}, we can obtain,

\begin{equation}\label{expand4}
\begin{split}
\widetilde{\Omega}(T,c_n)&\simeq\widetilde{\Omega}(T,N)\\
&\simeq \frac{2a_0}{\lambda^2N_1^2}(T/T_c-1)N^2+\frac{-16\widetilde{J}_fa_1}{\lambda^6N_1^4}N^4.
\end{split}
\end{equation}
We can see that this is just the Ginzburg-Landau (GL) theory of ferromagnetic model.

Now the grand thermodynamic potentials of the holographic model  in the gravity side and in the boundary theory side are in hand, we can use the AdS/CFT duality to relate them, which gives,
\begin{equation}\label{AdSCFTid}
\Omega_{QFT}=\Omega_{\text{gravity}}.
\end{equation}
Here we should mention that the dual theory in the boundary being an Ising-like model is an assumption. With this equality,
At the critical temperature, comparing \eqref{expand4} with \eqref{Veffs1}, we can obtain,
\begin{equation}\label{identiNS}
\frac{-16\widetilde{J}_fa_1}{\lambda^6N_1^4}N^4\sim\lambda_s\widetilde{s}^4_{cl}
\end{equation}
Note that in materials, the spontaneous magnetization is  proportional to the expectation value of $z$-component of spin, i.e., $N\propto\widetilde{s}_{cl}$. Following the definition of $\widetilde{J}_f$ in Eqs.~\eqref{rhop} and $\lambda_s$ in Eqs.~\eqref{massS}, we see that if $J=0$, then  Eq.~\eqref{identiNS} leads to $K=0$.  Therefore we can interpret the parameter $J$ in  action~\eqref{1} as  the deviation from the standard Ising model in the boundary theory.

Here some remarks are in order. First, the equality Eq.~\eqref{AdSCFTid} is exact according to the AdS/CFT correspondence. However, the expressions for the grand thermodynamic potentials in both sides are just some approximations. In gravity side, because we used the probe limit which neglects the back reaction of polarization field to the background geometry and to external field and we only computed the grand thermodynamic potential in the classical level which neglects the quantum correction. In the boundary side, we used mean field expansion and only took  the tree level of the quantum fluctuation into account. So the relationship \eqref{identiNS} is only approximately valid and our explanation for $J$ is only a qualitative description. Second, the Ising-like model~\eqref{Ising1} can only describe the local moment system, which is usually suitable for insulator. For metal magnetic materials, we need to use different model. So what the meaning of $J$ is  in these materials needs to be considered in the future.

 With the  grand thermodynamic potential in Eq.~\eqref{expand4}, we can obtain the expression of magnetic moment in the ferromagnetic phase as,
\begin{equation}\label{onshellN}
N/\lambda^2=\sqrt{\frac{N_1^2a_0}{-16\widetilde{J}_fa_1}}(1-T/T_c)^{1/2}.
\end{equation}
This confirmed the critical behavior obtained in the numerical calculations and the critical exponent $1/2$ is an exact result.  We can compute  all the coefficients  appearing in \eqref{onshellN} and compare them with the numerical ones. We  can also compute the  grand thermodynamic potential when magnetic field is nonzero and get the magnetic susceptibility and hysteresis loop. The details will be given in  Appendix~\ref{app2}.

\section{Probe limit by neglecting the back reaction of all matter fields}
\label{prob2}
\subsection{Spontaneous magnetization and susceptibility}
In the probe limit by neglecting the back reaction of  all matter fields including the Maxwell
field,  the  background geometry is just the planar AdS Schwarzschild  black brane with,
\begin{eqnarray}\label{32}
f(r)&=&1-\frac{r_h^3}{r^3},~~~~~~T=\frac{3r_h}{4 \pi}.
\end{eqnarray}
In the background, the equations for Maxwell field  read,
\begin{eqnarray}\label{33}
(m^2-\frac{4 J \rho^2}{r^4})p-\phi'&=&0, \nonumber \\
\phi''+\frac{2}{r}\phi'-\lambda^2(\frac{p'}{2}+\frac{p}{2r})&=&0.
\end{eqnarray}
We  can directly obtain $\phi(r)=\int_{r_h}^{\infty} dr(m^2-4 J\rho^2/r^4)p$ with $\phi(r_h)=0$
and put it into the equation of $\phi''$,
then the equation for $p$ can be obtained.
So in this probe limit, we need only solve the three  equations of matter fields numerically,
\begin{eqnarray}\label{34}
\rho''+\frac{f'}{f}\rho'-\frac{m^2+4 J p^2}{r^2 f}\rho+\frac{B}{r^2 f}&=&0,\nonumber\\
(m^2-\frac{4 J \rho^2}{r^4})p-\phi'&=&0,\\
p'+\left(\frac2r+\frac{32J \rho \rho'}{\lambda^2 r^4+16 J \rho^2-4 m^2r^4}\right)p&=&0.\nonumber
\end{eqnarray}
Note that in this case, there does not exist the $AdS_2$ geometry, the near horizon geometry of an extremal black brane with vanishing temperature.
To have the spontaneous magnetization when the temperature is lowered,  the restrictions \eqref{20}  need to be reconsidered. To find the
restriction about the parameters, let us consider  Eqs.~\eqref{34} in the high temperature region where $\rho$ vanishes and  the solutions for $\phi$ and $p$ are the same expressions shown
in Eqs.~\eqref{16}.  We can read off  the effective mass square of $\rho$ at the horizon as,
\begin{equation}\label{effm2}
m^2_{\rho\text{eff}}=m^2+4Jp(r_h)^2=m^2+\frac{4J\mu^2}{m^4r_h^2}=m^2+\frac{9J\mu^2}{m^44\pi^2 T^2}.
\end{equation}
Because of $J<0$, the temperature term contributes a negative term into the effective mass square, which is divergent when $T\rightarrow0$.
Thus  we see that in the grand canonical ensemble, the instability always appears provided that the temperature is low enough.
As a result, in this case,  we need not the restrictions in \eqref{20}. But, the parameters  have to satisfy the condition  $\lambda^2<4 m^2$.
To see this, let us  recall the charge density in Eq.~\eqref{freeE1}. In this probe limit, we have,
\begin{equation}\label{charge2}
\sigma(T,\mu)=2 (1-\frac{\lambda^2}{4 m^2})\frac{4\pi T\mu}3.
\end{equation}
Then we reach,
\begin{equation}\label{charge3}
\left(\frac{\partial\sigma}{\partial\mu}\right)_T=2 (1-\frac{\lambda^2}{4 m^2})\frac{4\pi T}3.
\end{equation}
We see that it is positive only when $\lambda^2<4 m^2$.

Now we consider the spontaneous magnetization in this probe limit. First, let us  compute the critical temperature $T_{c}$ when $B=0$.
Similar to the case in the first kind of  probe limit,
the polarization field $\rho$ is a small quantity near the critical temperature,
we can neglect the nonlinear terms of $\rho$ in  the equations of $\phi$ and $p$.
Then we can get $\phi(r)=\mu(1-r_h/r)$ and $p(r)=\mu r_h/r^2 m^2$ which are identical with
expressions in Eqs.~\eqref{16}. The equation of $\rho$  is just Eq.~\eqref{18}.
At the horizon, the initial conditions are,
\begin{eqnarray}\label{35}
\rho'&=&\frac{m^6+4 J \mu^2}{3 m^4},~~~~~\rho(r_{h})=1.
\end{eqnarray}
When we perform the numerical computation, we can first fix the horizon radius $r_h=1$ so the temperature is also fixed. By adjusting the chemical potential, we shoot the boundary condition $\rho_+=0$. In the end, we can use the scaling transformations \eqref{10} and \eqref{11} to transform our results into the case in grand canonical ensemble where the chemical
potential is fixed. As a typical example, we also choose parameters as $m^2=-J=1/8$ and $\lambda=1/2$.
The critical temperature is $T_{c}/\mu\simeq1.7871$.
Similarly, we can plot the relationship between $\rho_{+}$ and shooting parameter $\mu$,
in order to examine whether $\rho$ gets  spontaneous condensation when $T<T_{c}$.
We  find that the solution of source free always appears,
which results in the spontaneous magnetization of the system, and  breaks the time reversal symmetry
in low temperatures.


\begin{figure}
\centering
\includegraphics[width=0.3\textwidth]{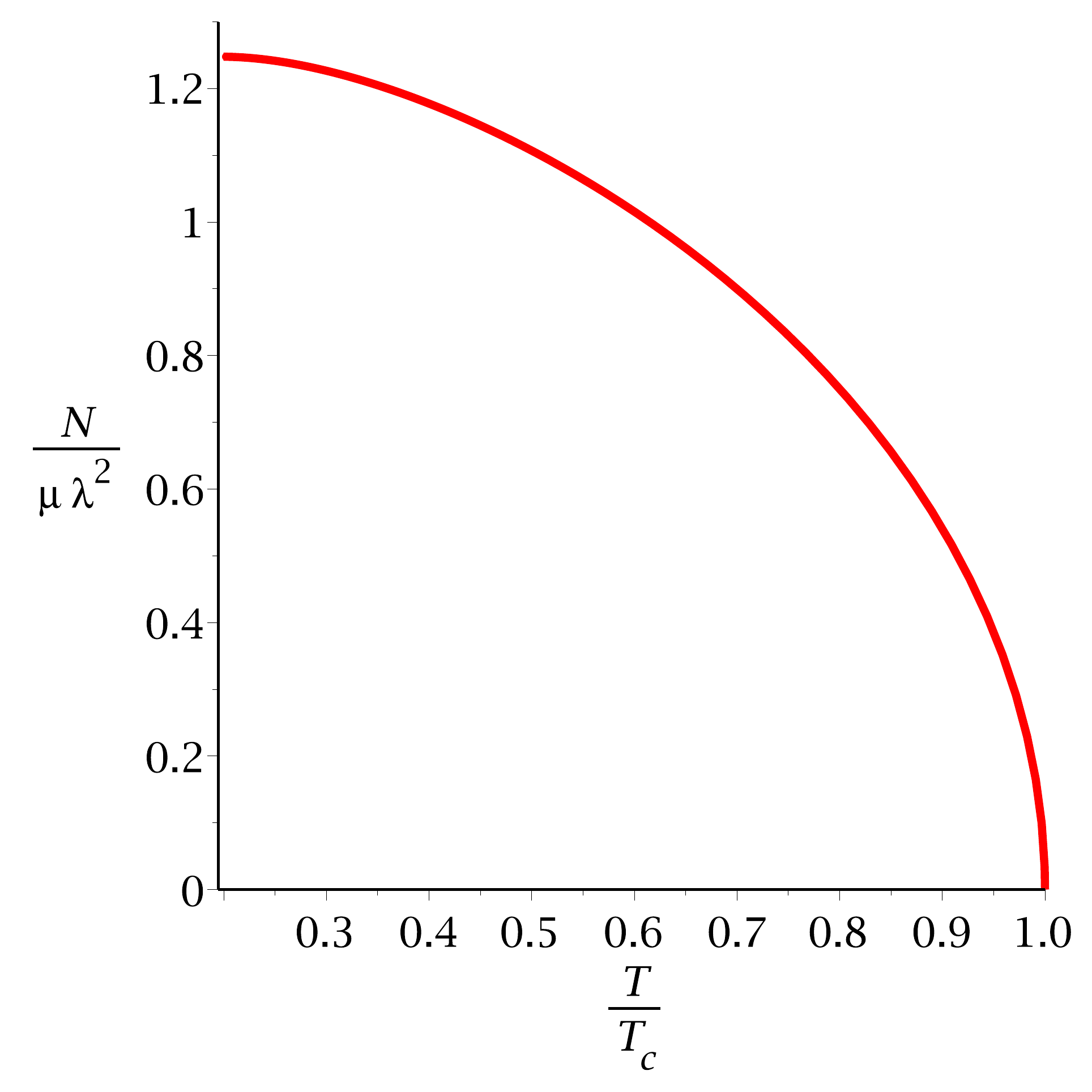}
\includegraphics[width=0.35\textwidth]{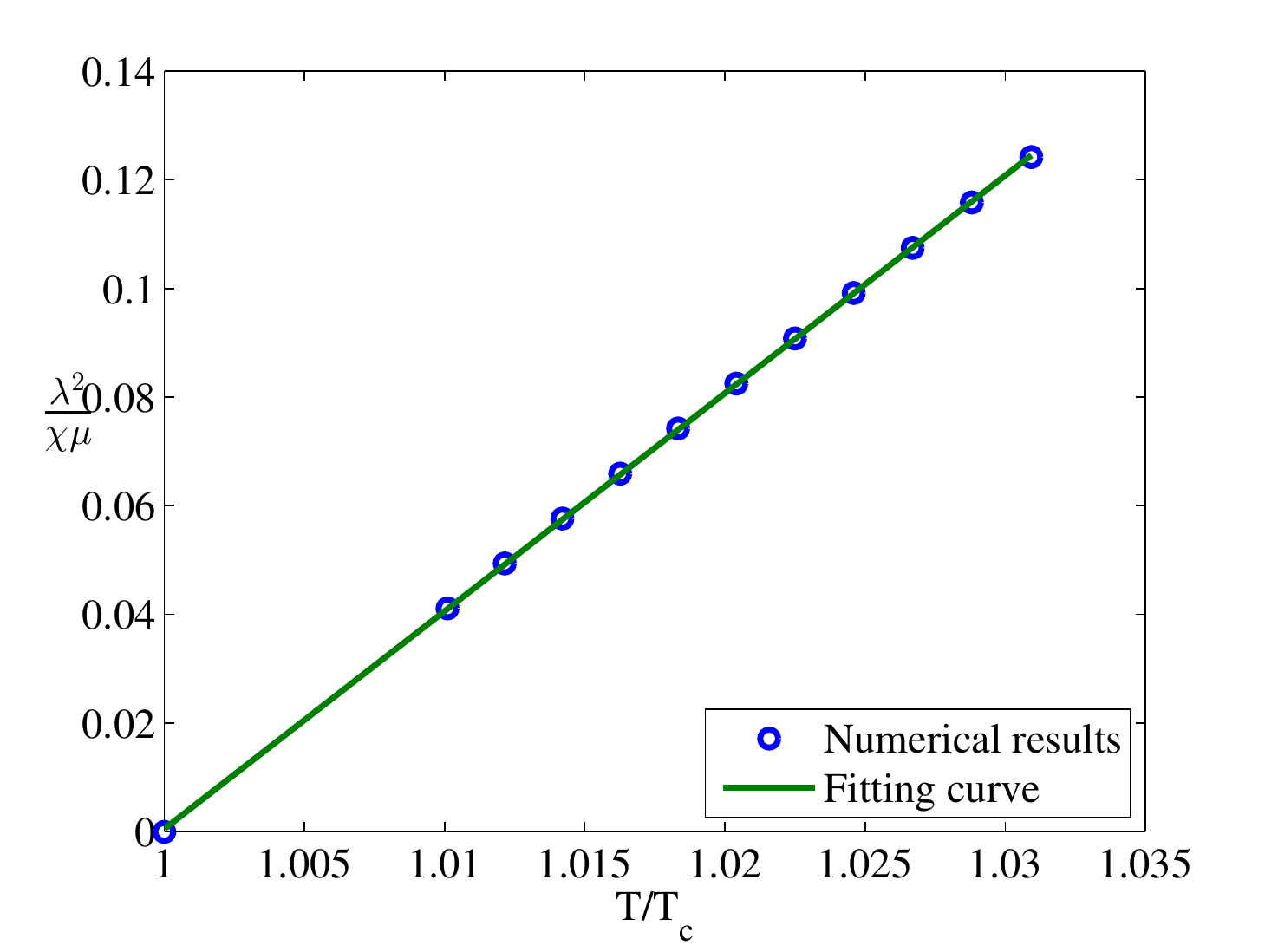}
\caption{Top panel: The magnetic moment $N$ as a function of temperature. Bottom panel: The behavior of the inverse
susceptibility density in the paramagnetic phase near the critical temperature.
Here, we choose parameters as $m^2=-J=1/8$. The temperature $T_{c}/\mu\simeq1.7871$.} \label{F3}
\end{figure}

When the temperature is  lower than the critical one  $T_{c}$,
we have to solve  Eq.~\eqref{34} to get the solution of the order parameter $\rho$,
and then compute the value of magnetic moment $N$,
which is also defined by Eq.~\eqref{defN1} with $a=0$.
In the top panel in Fig.~\ref{F3}, we plot the value of magnetic moment $N$ as a function of temperature.
We see that when the temperature is lower than $T_{c}$, the non-trivial solution $\rho\neq0$ and
spontaneous magnetic moment  appears.
It corresponds a time reversal symmetry breaking spontaneously.
In addition, let us stress here that if the boundary spatial dimension is three,
the spatial rotational symmetry is also broken spontaneously,
since a nonvanishing magnetic moment chooses a direction as a special.
The numerical results show that this phase transition is a second order one with the behavior
$N\propto\sqrt{1-T/T_{c}}$ near the critical temperature.
The result is still consistent with the mean field theory description of
the paramagnetism/ferromagnetism phase transition.

Next we calculate the static magnetic susceptibility in this probe limit,
defined by Eq.~\eqref{28}.
Based on the previous analysis, the magnetic susceptibility is still obtained by solving Eq.~\eqref{29},
and the only difference is the form of $f(r)$.
Thus we can also  set the magnetic field $B=1$ and   get $\frac{\lambda^{2}}{\chi \sqrt{\widetilde{\mu}}}=5.7(T/T_{c}-1)$, which
satisfies the Curie-Weiss law of ferromagnetism in the region of $T>T_{c}$.    Its inverse is shown in  the bottom panel of Fig.~\ref{F3}.

\subsection{DC conductivity in the ferromagnetic phase}
The electric transport is also an important property in the materials involving spontaneous magnetization. Now let us study how  the DC conductivity is influenced by spontaneous magnetization in this model. In order to simplify our computation in technology, we will work in the probe limit by neglecting back reactions of all the matter fields. This limit can give out the main features near the critical temperature. However, in the case of near zero temperature, we have to consider the model with full back reaction. We will consider this in the future.

 To compute the conductivity, we have to consider  some perturbations for gauge field with harmonically time varying electric field.
Due to the planar symmetry at the boundary, the conductivity is isotropic.  Thus for simplicity, we just compute the conductivity along the $x$-direction.
According to the dictionary of AdS/CFT,  we consider the perturbation  $\delta A_x=\epsilon a_x(r)e^{-i\omega t}$. In the probe limit,
this perturbation will also lead to the perturbations of polarization field in the first order of $\epsilon$. As a result,
we have to consider the perturbations for all the components of gauge field and polarization field. However, if we only care the conductivity in the low frequency limit,
i.e., $T\gg\omega\rightarrow0$, the problem can be simplified. In the low frequency limit, we only need turn on the three perturbations,
\begin{equation}\label{pert1}
\begin{split}
\delta A_x&=\epsilon a_x(r)e^{-i\omega t},\\
M_{rx}&=\epsilon C_{rx}(r)e^{-i\omega t},\\
M_{ty}&=\epsilon C_{ty}(r)e^{-i\omega t},
\end{split}
\end{equation}
and  corresponding equations for the three perturbations in the low frequency limit read
\begin{subequations}\label{pert2}
\begin{align}
C_{ty}''-\frac{m^2C_{ty}}{r^2f}-\frac{4Jp\rho C_{rx}}{r^2}+O(\omega)=0,\label{pert2a}\\
C_{rx}-\frac{a_x'}{m^2}-\frac{4Jp\rho C_{ty}}{r^4f m^2}+O(\omega)=0,\label{pert2b}\\
[r^2f(a_x'-\lambda^2C_{rx}/4)]'+\frac{a_x\omega^2}{r^2f}+O(\omega)=0,\label{pert2c}
\end{align}
\end{subequations}
with $p$ and $\rho$ determined by Eqs.~\eqref{34}. Here $O(\omega)$ is the terms with order of $\omega$ which can be neglected when $\omega\rightarrow0$. In general, the term  $\omega^2/r^2f(r)$ can not be neglected since $f(r)$ is zero at the horizon, which leads to that the limit of $\omega\rightarrow0$ is ambiguous. However, at the horizon, if we impose the ingoing conditions for $C_{rx}, C_{ty}$ and $a_x$,
\begin{equation}\label{init4}
\begin{aligned}
C_{ty}&=C_{ty}^{(0)}+C_{ty}^{(0)}(r-r_h)+\cdots,\\
C_{rx}&=e^{-i\omega r_*}[C_{rx}^{(0)}+C_{rx}^{(1)}(r-r_h)+\cdots], \\
a_x&=e^{-i\omega r_*}[a_{x}^{(0)}+a_{x}^{(1)}(r-r_h)+\cdots]
\end{aligned}
\end{equation}
with $r_*=\int dr/(r^2f)$,  we find the system has a well-defined limit  when $\omega\rightarrow0$ if $T\neq0$.
At the AdS boundary with the source free condition, we have the following asymptotic solutions,
\begin{equation}\label{pert3}
\begin{split}
C_{ty}&=C_{ty+}r^{(1+\delta)/2}+C_{ty-}r^{(1-\delta)/2}+\cdots,\\
C_{rx}&=-\frac{a_{x-}}{r^2 m^2}+\cdots,~a_x=a_{x+}+\frac{a_{x-}}{r}+\cdots.
\end{split}
\end{equation}
Then the gauge/gravity duality implies  that electric current $\langle J_x\rangle=a_{x-}$ and the DC conductivity is given by,
\begin{equation}\label{DCconduc}
\sigma=\lim_{\omega\rightarrow0}\frac{a_{x-}}{i\omega a_{x+}}.
\end{equation}

\begin{figure}
\begin{center}
\includegraphics[width=0.3\textwidth]{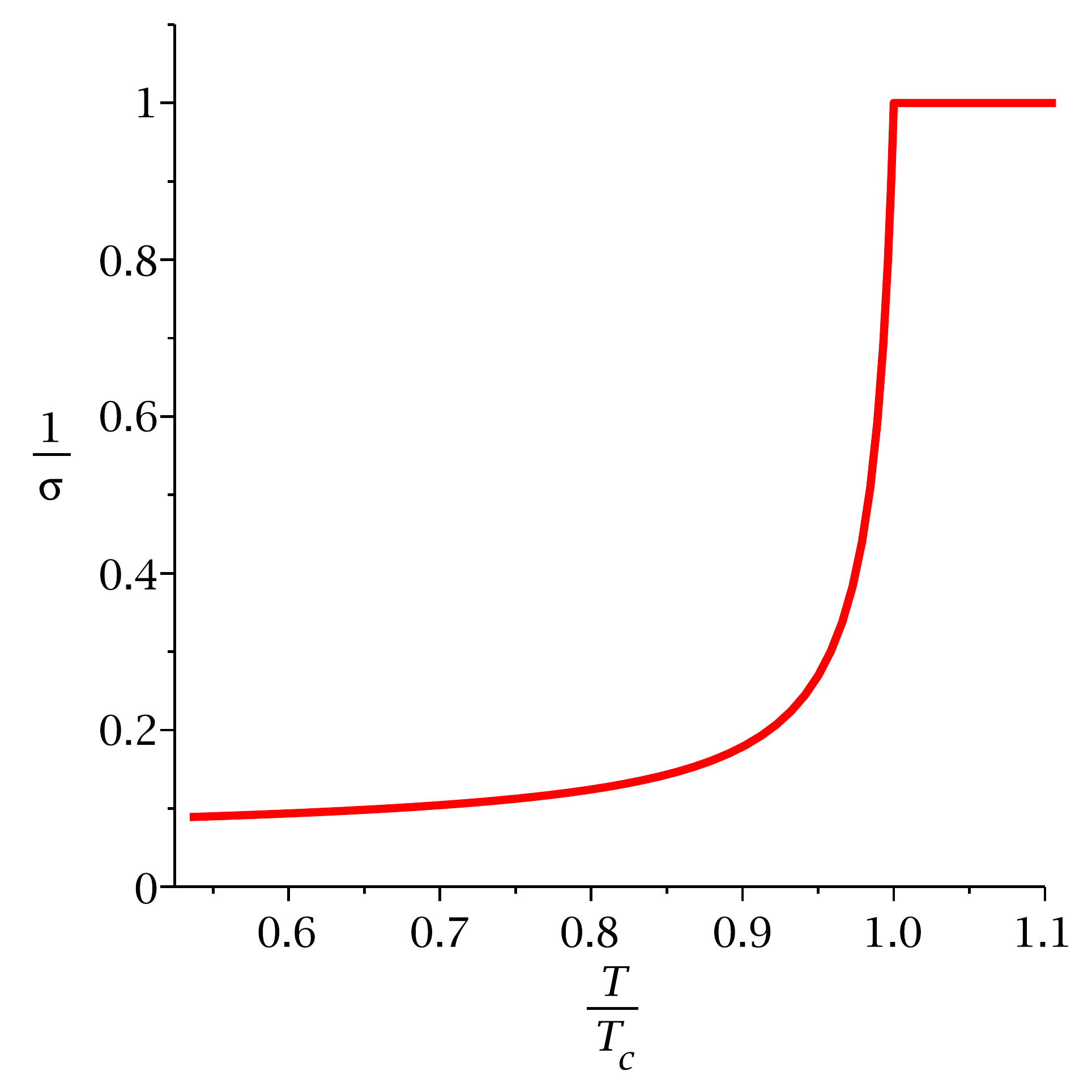}
\includegraphics[width=0.3\textwidth]{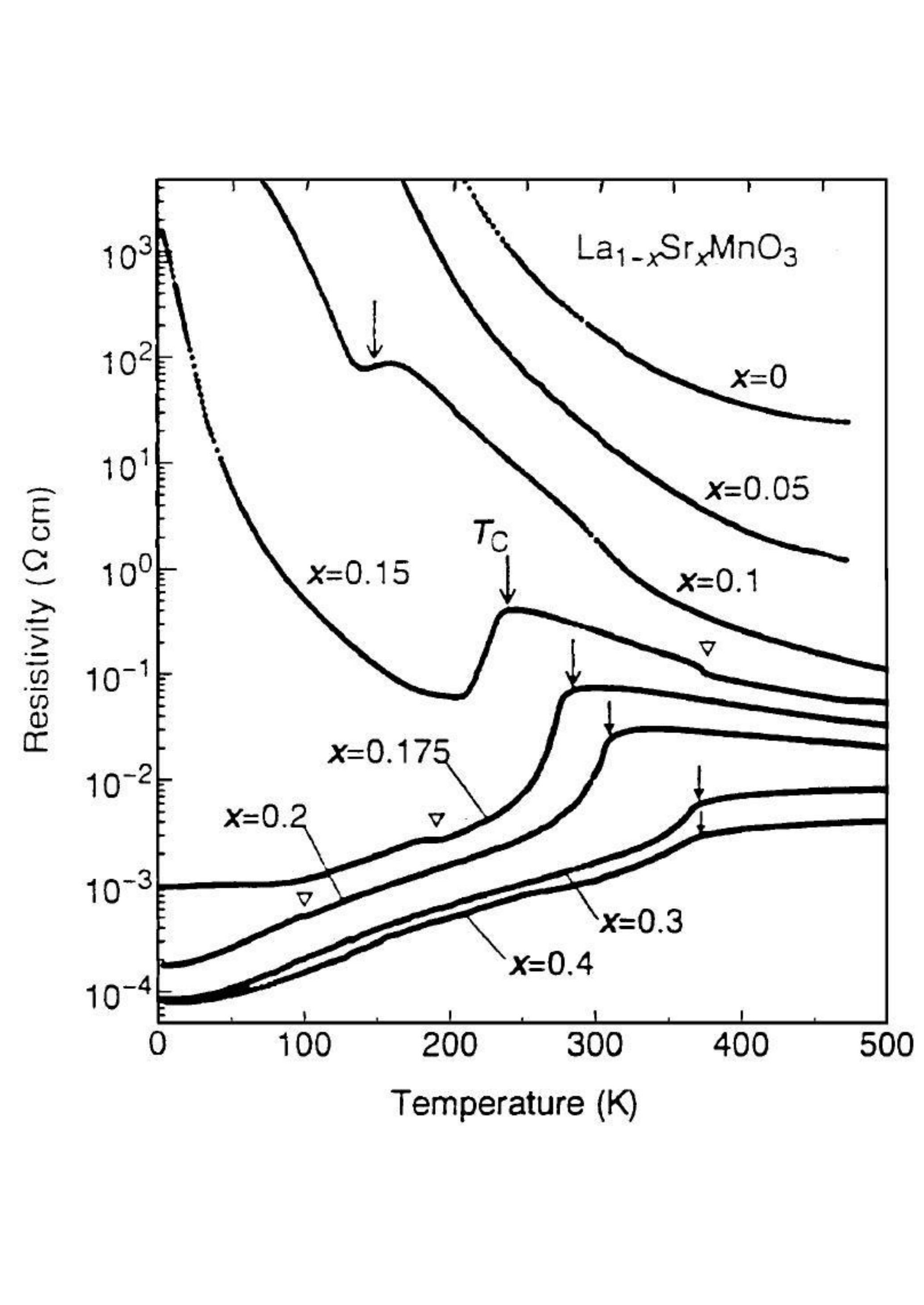}
\caption{Top panel: DC resistivity vs temperature in our model. Here we choose parameters as $m^2=-J=1/8$ and $\lambda=1/2$. The critical temperature $T_{c}/\mu\simeq1.7871$. Bottom panel: Temperature dependence of resistivity for various single crystals of La$_{1-x}$Sr$_x$MnO$_3$. Arrows indicate the
Curie temperature. For more details,  see Ref.~\cite{Urushibara}. }
\label{TO1}
\end{center}
\end{figure}

As a holographic application of the membrane paradigm of  black holes, we can direct obtain the DC conductivity from  Eqs.~\eqref{pert2} using the method proposed by Iqbal and Liu  in \cite{Iqbal:2008by}. In fact, the transport coefficients in the dual field theory can be obtained from the horizon geometry of the dual gravity in the low frequency limit. Applying this into $U(1)$ gauge field, this conclusion implies that the DC conductivity is given by the coefficient of the gauge field kinetic term evaluated at the horizon. To see this, we assume that $T>0$ and $\omega\rightarrow0$, then we can neglect all the terms of $\omega$ in Eqs.~\eqref{pert2}. We first note that,
\begin{equation}\label{pert4}
\begin{split}
&\lim_{r\rightarrow\infty}r^2 f(r)(a_x'-\lambda^2C_{rx}/4)=r^2\left(\frac{-\langle J_x\rangle}{r^2}+ \frac{\lambda^2\langle J_x\rangle}{4r^2m^2}\right)\\
&=-(1-\lambda^2/4m^2)\langle J_x\rangle.
\end{split}
\end{equation}
 Eq.~\eqref{pert2c} shows that this quantity is conserved along the direction $r$. So at the horizon, using Eqs.~\eqref{pert2b} and \eqref{pert2c}, we have,
\begin{equation}\label{pert5}
\begin{aligned}
&-(1-\lambda^2/4m^2)\langle J_x\rangle=\lim_{r\rightarrow1}\left.r^2f\left(a_x'-\lambda^2C_{rx}/4\right)\right|_{r=r_h},\\
&=r^2f\left[\left(1-\frac{\lambda^2}{4m^2}\right) a_x'+\frac{4Jp\rho C_{ty}}{m^2f}\right]_{r=r_h}.
\end{aligned}
\end{equation}
Combining Eqs.~\eqref{pert2a} and~\eqref{pert2b} and considering  the fact that $C_{ty}$ is regular at the horizon, we have,
\begin{equation}\label{pert6}
\left[m^2+\frac{16J^2p^2\rho^2}{m^2r^4}\right]C_{ty}=- \frac{4Jp\rho}{m^2}fa_x'
\end{equation}
at $r\rightarrow r_h^+$. Thus we have  from Eqs.~\eqref{pert6} and~\eqref{pert5} that
\begin{equation}\label{pertb1}
\begin{aligned}
&-(1-\lambda^2/4m^2)\langle J_x\rangle\\
&=\lim_{r\rightarrow r_h^{+}}r^2fa_x'\left[1-\frac{\lambda^2m^2}{4(m^4+16J^2p^2\rho^2/r^4)}\right].
\end{aligned}
\end{equation}
Now let us take the ingoing condition for $a_x$ at the horizon, which tells us that,
\begin{equation}\label{ingoing1}
r^2fa_x'=\frac{d}{dr_*}a_x=-i\omega a_x,~~\text{at}~r\rightarrow r_h^+,
\end{equation}
finally we get,
\begin{equation}\label{pertb2}
\langle J_x\rangle=\frac{i\omega a_x(r_h)}{1-\frac{\lambda^2}{4m^2}}\left[1-\frac{\lambda^2m^2}{4(m^4+16J^2p_0^2\rho_0^2/r_h^4)}\right].
\end{equation}
Here $p_0$ and $\rho_0$ are the initial values of $p(r)$ and $\rho(r)$ at the horizon, which can be computed from Eqs.~\eqref{34}.
In the low frequency limit, Eqs.~\eqref{pert2} imply that the electric field is constant, i.e., $\lim_{r=r_h}a_x(r)=a_{x +}$.
So we have seen that we can obtain the DC conductivity near the horizon as,
\begin{equation}\label{DCexp}
\sigma=\frac{1}{1-\lambda^2/4m^2}\left[1-\frac{\lambda^2m^2}{4(m^4+16J^2p_0^2\rho_0^2/r_h^4)}\right].
\end{equation}
In the top panel of Fig.~\ref{TO1}, we plot the numerical result on the  DC resistivity $1/\sigma$ as a function of temperature. We see that the resistivity shows a metallic behavior when the temperature
is  below the Curie temperature.

With the appearance of ferromagnetism, DC resistivity decreases when  the sample gets cooling, which shows in many interesting phenomena in condensed matter physics, especially in a class of manganese oxides which are widespread because of the discovery of colossal magnetoresistance (CMR) and  receive a lot of interest  both in theory and experiment~\cite{Dagottoa,Nagaev}. Note that this effect has a completely different physical origin from the ``giant" magnetoresistance observed in layered and clustered compounds. In recent twenty years, CMR is among the main topics of study within the area of strongly correlated electron systems and its popularity is reaching the  level comparable to that of the high-temperature superconducting cuprates. In the bottom panel of Fig.~\ref{TO1}, we show the experimental data from a typical CMR material La$_{1-x}$Sr$_x$MnO$_3$ as an example. We see that our model gives a very similar behavior as  the latter with $x\geq0.175$.  Of course, we should mention here that there still exist some differences between our model result and experimental data on CMR. In general, when $T>T_c$, the material shows a semiconductor or insulator behavior and the DC resistivity increases with cooling the sample. In our model, however,  the DC resistivity is a constant when $T>T_c$.  So this model only gives partial property of CMR when $T< T_c$.  But  this is  an exciting and enlightening result, because it implies that this model can lead to a possibility to build a holographic CMR model and to investigate this typical and important strong correlated electrons system in the AdS/CFT setup.  We are going to investigate this issue in the future.

\section{Phase transition with back reaction}
\label{backrec}

In previous sections, we have studied the spontaneous magnetization in two kinds of probe limit.  However,  probe limit may lead to  some information lost.  For example, in some holographic superconductor models, it will lead to the appearance of  the first or zeroth order phase transition when  the strength of back reaction gets beyond some values (see Refs.~\cite{Ammon:2009xh,Cai:2013aca,Cai:2014ija}, for example).  In addition, when the temperature is low enough, the probe limit may lose its validness. To have a complete phase diagram for the holographic model, we need to go beyond the probe approximation and to include the back reaction.

\subsection{On-shell free energy}

The model admits various solutions, in order to determine which phase is thermodynamically favored,
we should calculate the free energy of the system for both normal phase and condensed phase.
In gauge/gravity duality the grand potential $\Omega$ of the boundary thermal state is
identified with temperature $T$ times the on-shell bulk action with Euclidean signature.
Since we are considering  a stationary problem, the Euclidean action is related to the Minkowski one by a minus sign as
\begin{eqnarray}\label{36}
-S_{E}&=&\int d^{4} x \sqrt{-g} (\mathcal{R}+\frac{6}{L^{2}}+\mathcal{L}_{m}),
\end{eqnarray}
where $\mathcal{L}_{m}=-F^{\mu\nu} F_{\mu\nu}+\lambda^{2} L_{2}$ and $g$ is the determinant
of the metric. We first show that, when evaluated on a solution, this action reduces to a simple
surface term at the AdS boundary. From the symmetries of the solution ($5$) and ($6$), the $yy$ component
of the stress energy tensor only has a contribution from the terms proportional to the metric.
Thus, the gravitational filed equations imply that
\begin{eqnarray}\label{37}
\mathcal{R}+\frac{6}{L^{2}}&=&2 \mathcal{R}^{y}{}_{y}+2 T^{y}{}_{y}.
\end{eqnarray}
The Euclidean action is then
\begin{eqnarray}\label{38}
\frac{-S_{E}}{V_{2}}&=&\int dr [(2 r^{3} f e^{a/2}-\lambda^{2} f \rho \rho' e^{a/2})'\nonumber\\
&~&-\frac{B (4 B-\lambda^{2} \rho) e^{a/2}}{r^{2}}],
\end{eqnarray}
The surface term on the horizon vanishes since $f(r_{h})=0$.
Thus we can get  the Euclidean action that contains the surface term at $r_{\infty}$ as,
\begin{eqnarray}\label{39}
\frac{-S_{E}}{V_{2}}&=&2 r^{3} f e^{a/2}-\lambda^{2} f \rho \rho' e^{a/2}|_{r=r_{\infty}}\nonumber\\
&~&-\int dr \frac{B (4 B-\lambda^{2} \rho) e^{a/2}}{r^{2}}.
\end{eqnarray}
As the first item of Eq.~\eqref{39} diverges when $r\rightarrow\infty$ and must be regulated.
This counter terms we need to regulate the action are the standard ones (see for example~\cite{31}):
\begin{eqnarray}\label{40}
S_{c.t.}&=&\frac{1}{2 \kappa^{2}}\int dx^3 \sqrt{-g_{\infty}} (-2 K+4/L)\mid_{r=r_{\infty}},
\end{eqnarray}
where $g_{\infty}$ is the induced metric on the boundary $r=r_{\infty}$ and $K=g^{\mu\nu}_{\infty}\nabla_{\mu} n_{\nu}$ is the trace of
the extrinsic curvature ($n^{\mu}$
is the outward pointing unit normal vector to the boundary).
The summation  $S_{\text{Euclidean}}^{\text{on-shell}}=S_{E}+S_{c.t.}$ is now finite in the limit $r\rightarrow \infty$.
The regularized action becomes, after considering the asymptotical forms in \eqref{13},
\begin{eqnarray}\label{41}
\Omega&=&T S_{\text{Euclidean}}^{\text{on-shell}}/V_2\nonumber\\
&=&-\int^{\infty}_{r_h} dr \frac{B(4 B-\lambda^{2} \rho) e^{a/2}}{r^2}+2 f_{0}.
\end{eqnarray}

\subsection{Tensor hairy solutions and phase transition}


We are interesting in the black brane solutions with nontrivial space-space component $\rho$ of the tensor field.
For this, we have to adopt the numerical method to find such solutions.
Without loss of generality, the location of $r_{h}$ can be fixed to be unity  in our numerical calculation.
We are then left with two independent parameters $\{\rho(r_{h}), p(r_{h})\}$.
By choosing $p(r_{h})$ as the shooting parameter to match the source free condition at $r\to \infty$, i.e.,
$\rho_{+}=0$, we finally have a one-parameter family of solutions labeled by the value of $\rho(r_{h})$.
Other coefficients can be expressed in terms of those parameters.
After solving the set of equations,
we can calculate  the spontaneous magnetization $N$ and free energy density.

\begin{figure}
\centering
\includegraphics[width=0.23\textwidth]{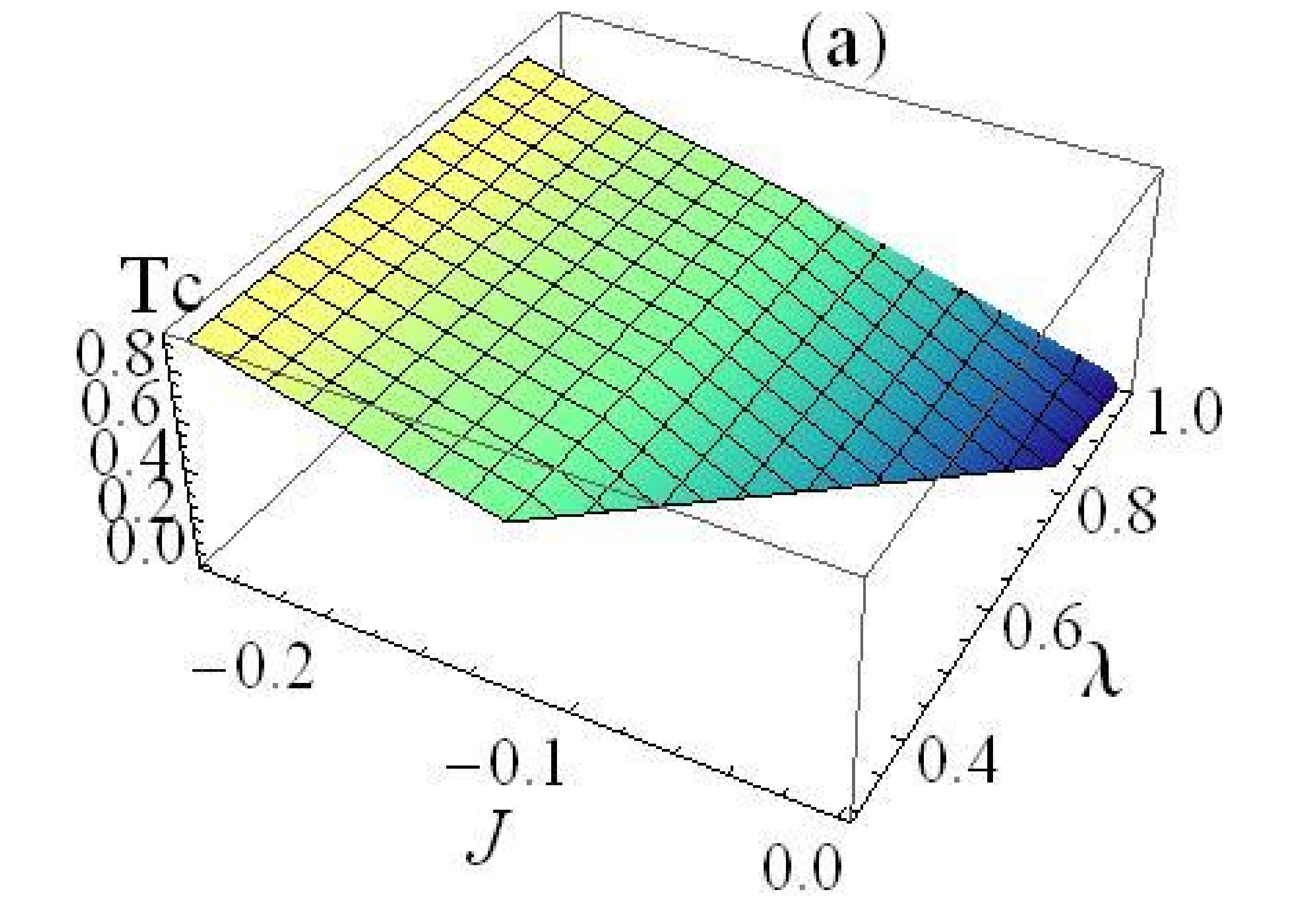}
\includegraphics[width=0.23\textwidth]{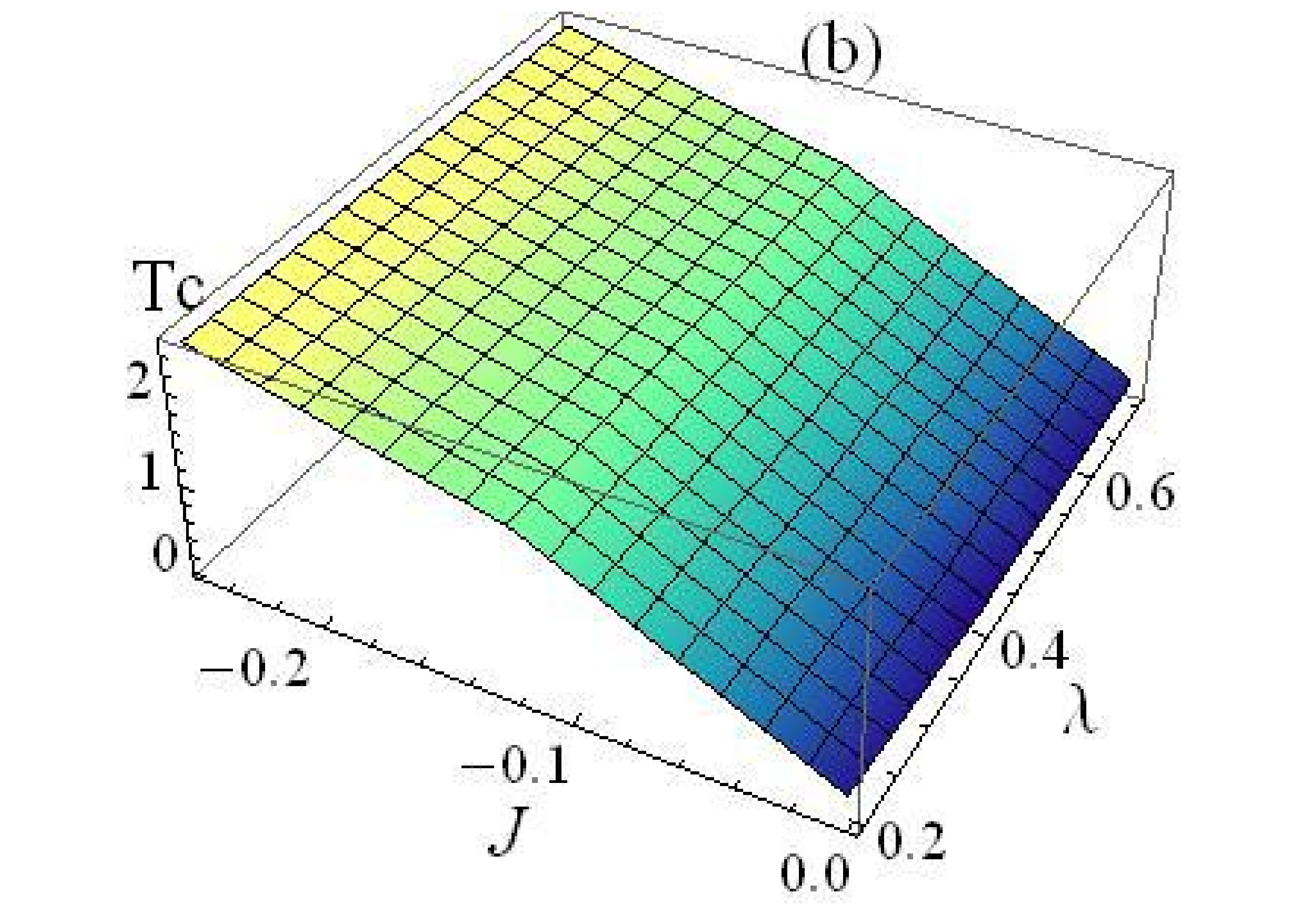}
\includegraphics[width=0.23\textwidth]{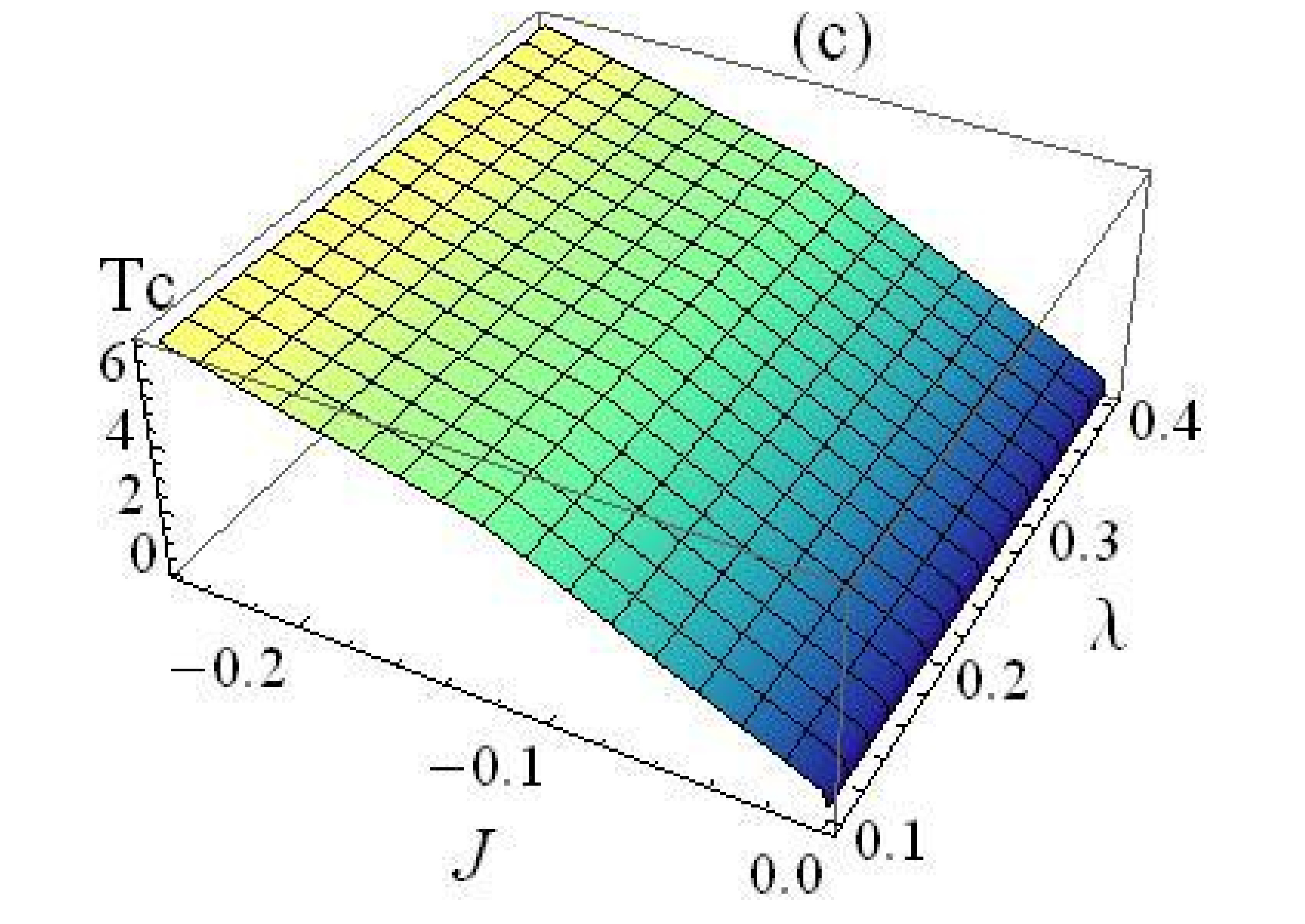}
\includegraphics[width=0.23\textwidth]{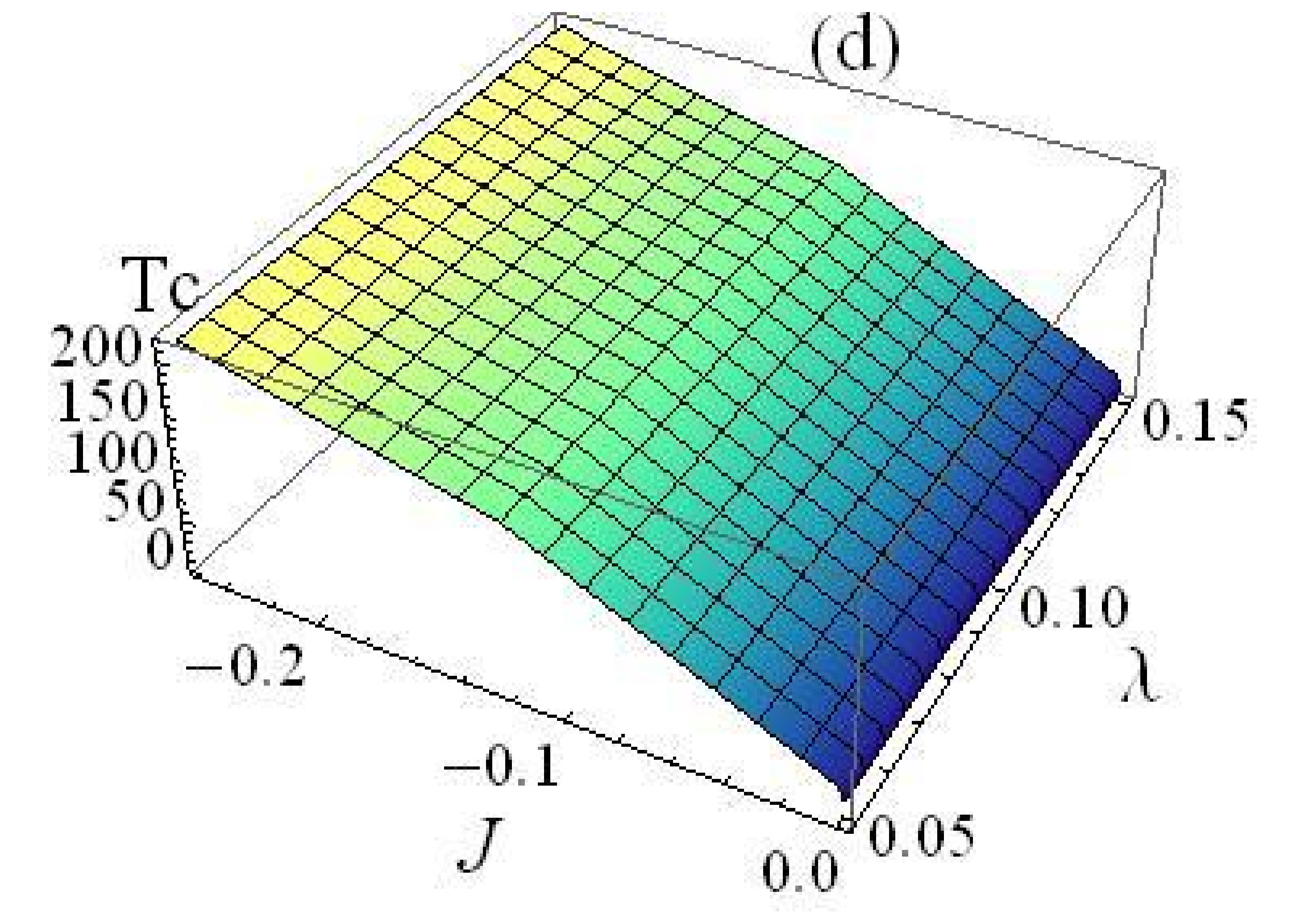}
\caption{The  critical temperature $T_{c}$ with respect to the model parameters $J$ and $\lambda$. Here (a), (b), (c) and (d) correspond to $m^{2}=1/4, 1/8, 1/16$, and $1/160$, respectively. In (a), some part of the surface is cut  because that $J$ and $\lambda$ should satisfy the relation \eqref{20}.}
\label{TclambdaJ}
\end{figure}

\begin{figure}
\centering
\includegraphics[width=170pt,height=130pt]{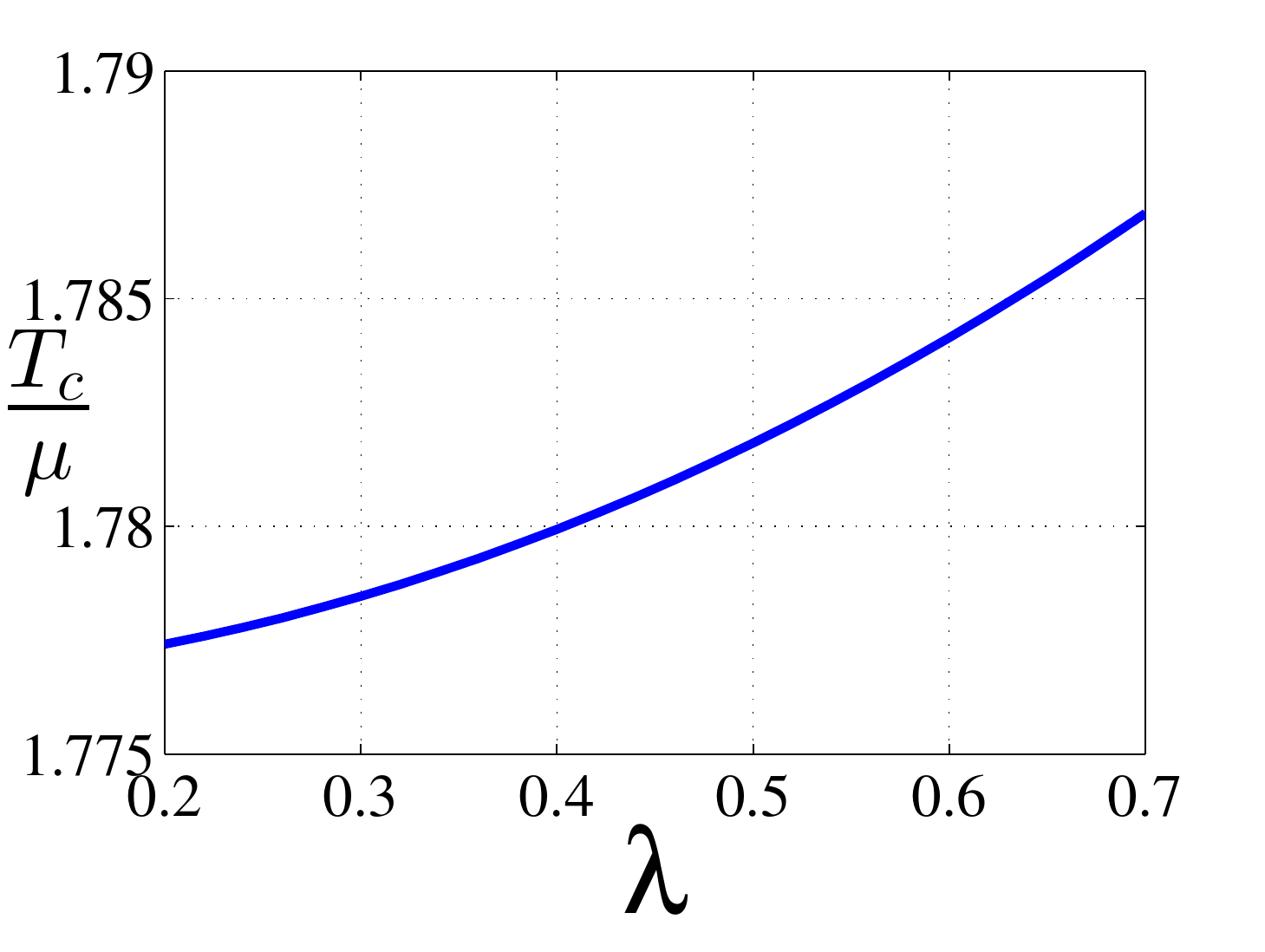}
\caption{The critical temperature $T_{c}$ as a function of $\lambda$ in the case with model parameter $m^2=-J=1/8$.}\label{Tclambda}
\end{figure}

  With a fixed $m^2$, we scan a wide range of $J$ and $\lambda$  within the limitation Eq.\eqref{20} in 3-dimension plane
in order to trace out the evolution of critical temperature $T_{c}$ versus these parameters.
Fig.~\ref{TclambdaJ} plots the  critical temperature $T_{c}$  as a function of  $J$ and $\lambda$ in the cases with $m=1/4$, $1/8$. $1/16$ and $1/160$, respectively.
We can see  from Fig.~\ref{Tclambda}  that  the critical temperature is weakly dependent of the parameter $\lambda$. Moreover, the smaller the value of $m^2$,
the larger the phase transition temperature $T_{c}$ when  the same value of $J$ and $\lambda$ are considered.
 For each value of
$\lambda$, the analytical black brane solution~\eqref{16} always exists.
However, for sufficiently low temperature, we always find additional solutions with
non-vanishing $\rho$ which are thermodynamically favored. That is to say, for each value of $\lambda$
we take, there is a phase transition occurring at a certain temperature $T_{c}$ where the
black brane developing a new ``tensor hair" with nontrival $\rho$ becomes thermodynamically favored. In the dual field
theory side, it means that magnetic moment acquires a vacuum expectation value
breaking the time reversal symmetry spontaneously.

%
%


\begin{figure}
\centering
\includegraphics[width=0.3\textwidth]{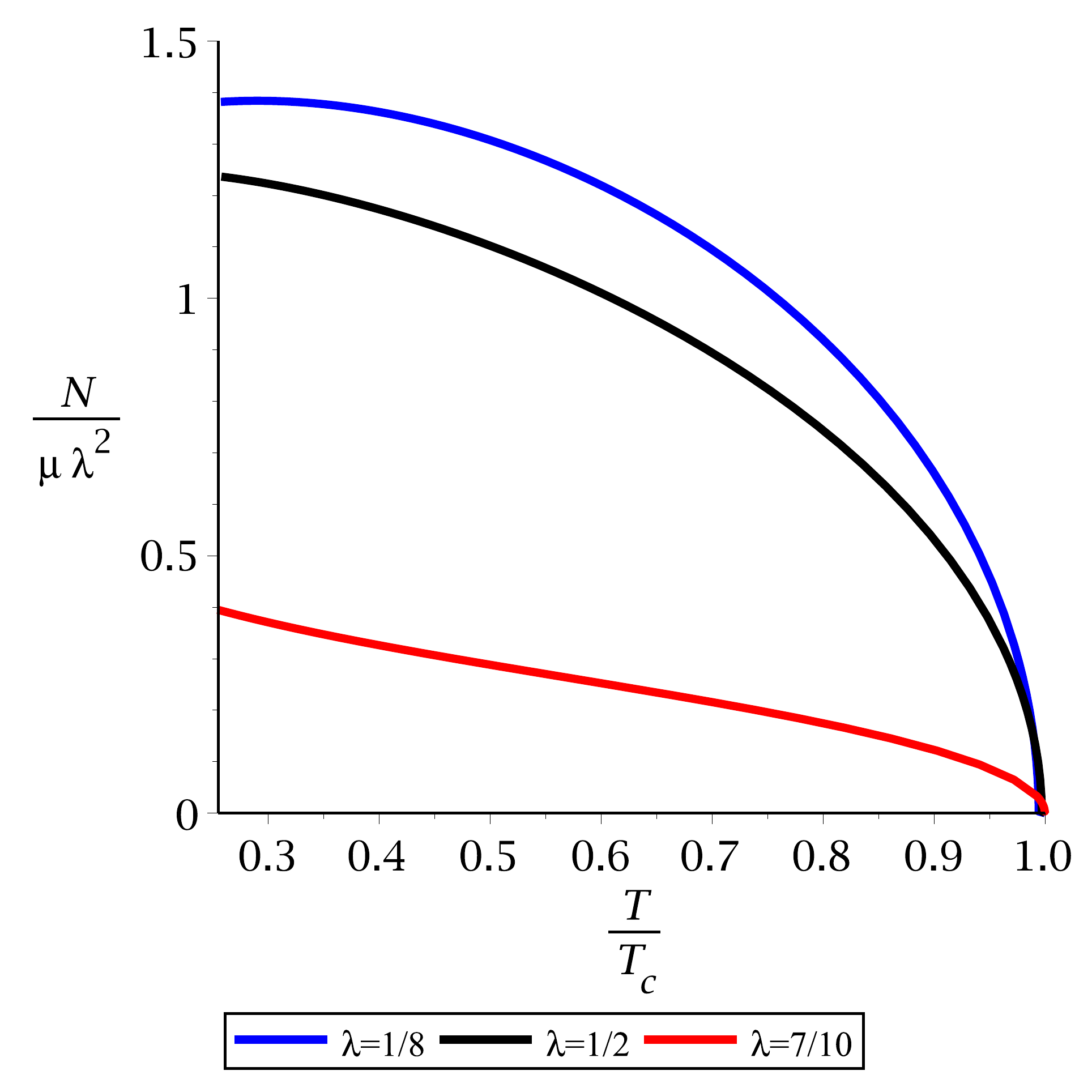}
\includegraphics[width=0.35\textwidth]{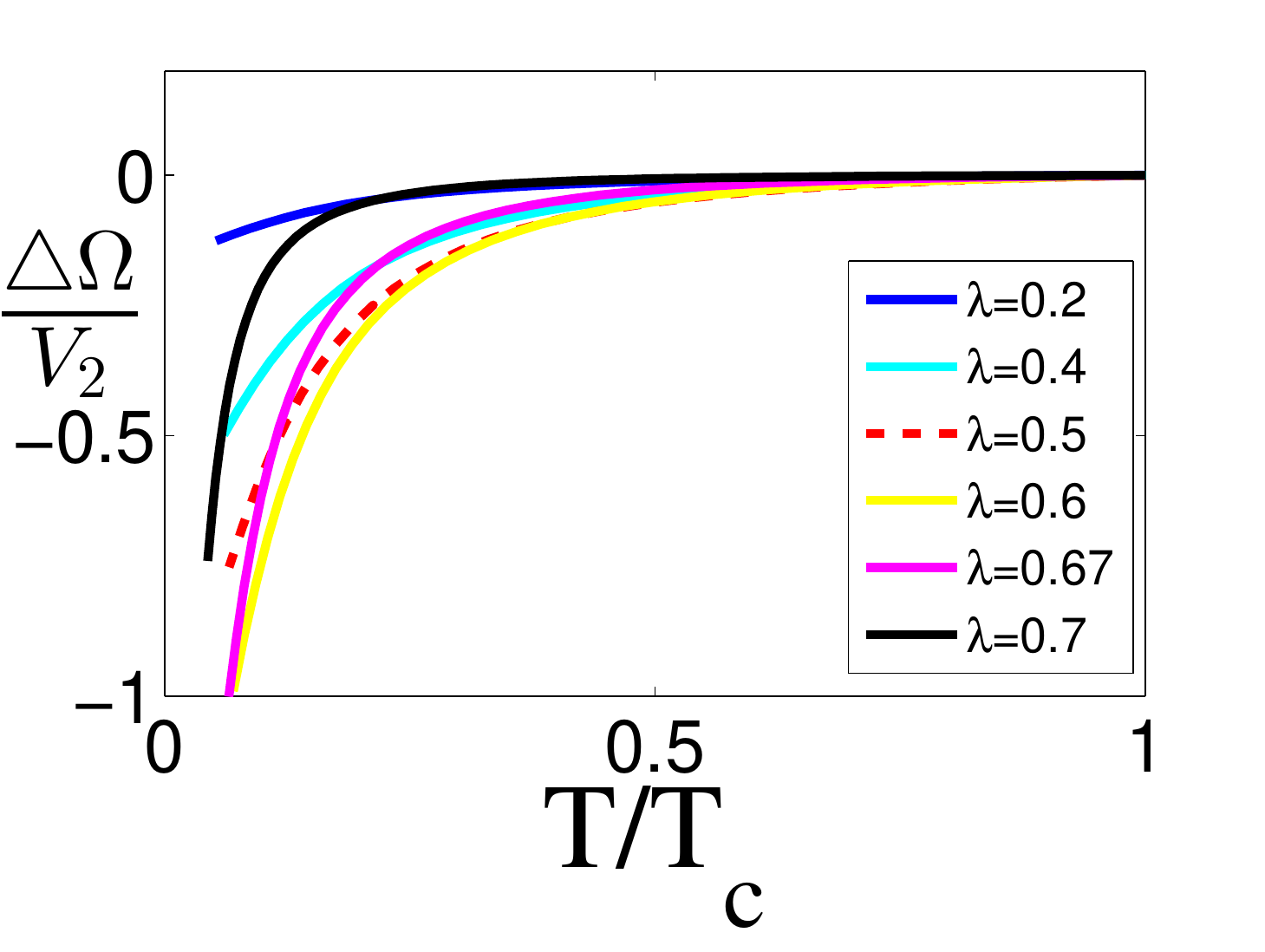}
\caption{The magnetic moment $N$ (top panel) and the grand potential deference $\Delta\Omega$ (bottom panel) as a function of temperature
with different $\lambda$.  Here $m^2=-J=1/8$ for both figures.} \label{F4}
\end{figure}

The top panel of Fig.~\ref{F4} presents the
condensate as a function of temperature for $\lambda=1/8, 1/2$ and $7/10$,
from which one can see that $N$ rises continuously from zero at $T_{c}$.
For small $\lambda$, the curve is  similar to the case  with the probe limit (compare the case of $\lambda=1/8$ in Fig.~\ref{F4} and  the plot in the bottom panel of Fig.~\ref{F1}),
We can see from the figure that when $\lambda$ gets larger, the condensate increases.  However, near the critical temperature,
  the square root behavior still holds as
\begin{equation}\label{NTsquare}
N\propto\sqrt{T-T_c}.
\end{equation}
The free energy difference of the condensed phase  and the normal phase
is expressed by $\Delta\Omega$, which is plotted  in the bottom panel of Fig.~\ref{F4}.
It is obvious that below the critical temperature $T_{c}$, the state with non-vanishing magnetic
``tensor hair'' is indeed thermodynamically favored over the normal phase because $\Delta\Omega$
always be less than zero.
Moreover, our numerical calculation indicates that the order of the phase transition is only
second order, no matter how we increase the strength of the back reaction  in parameters allowed region, i.e., $\lambda^2/4m^2<1$.
The first order or even the zero order phase transition does not appear in this model.

\section{Summary and discussion}
\label{summ}
In this paper we have presented a holographic model to realize the paramagnetism/ferromagnetism
phase transition in  AdS black brane background
by introducing a massive 2-form field in the bulk.
This 2-form field couples to the background Maxwell field strength and carries self interaction.

The model admits  a new analytical black brane solution with  a non-trivial time-space component
of the tensor field. The properties of the black brane solution depend on the value of $\lambda^2/4m^2$. When $\lambda^2/4m^2<1$,  this new solution is very similar to the planar AdS RN black hole. But if  $\lambda^2/4m^2>1$, the black brane solution is chemically  unstable in grand canonical ensemble unless the temperature and chemical potential  satisfy  some additional conditions. In that case, there is not  corresponding extreme black hole, i.e., the horizon radius is zero when temperature goes to zero. A very special case is that when $\lambda^2/4m^2=1$,  the space-time geometry is just the planar AdS Schwarzschild  geometry but both the U(1) field and 2-form field nonvanish.

For our goal in this paper  we focused on the case with  $\lambda^2/4m^2<1$ so that an asymptotic AdS$_2$ geometry near the horizon emerges when the  temperature tends to zero. By this emergent AdS$_2$ geometry, we obtained the condition in which the spontaneously symmetry breaking can happen. If the parameters satisfy the condition, time reversal symmetry will be broken spontaneously and the paramagnetism/ferromagnetism phase transition can happen when the temperature is lower than a critical value.

In order to understand the properties of this holographic ferromagnetic phase transition, we investigated the paramagnetism/ferromagnetism phase transition
in two kinds of probe limit and in the case with  full back reaction, respectively.

In the case of the first  kind of probe limit where the model parameters $\lambda\rightarrow 0$. This probe limit neglects the back reactions of the 2-form field to the Maxwell  field and  background geometry and makes it simple to study the behavior of spontaneous magnetization. In this probe limit, we computed the critical exponents by both numerical and analytical approaches, which are agreement with the mean field results. In addition,  we obtained a Ginzburg-Landau-like free energy  near the critical temperature for the holographic model and argued that it is agreement with results from the Ising-like Hamiltonian in film system of the condensed matter physics and related the model parameter $J$ to the parameters in the Ising-like model.

In the second kind of probe limit, we neglected all back reaction of matter fields to the background geometry but considered the interaction between the tensor field  and the Maxwell field.  In this probe limit, we are able to  study the influence of spontaneous magnetization on the electric transport properties in a relatively simple way, where the background geometry is fixed. We found that the critical exponents are the same as ones in the first kind of probe limit. We also computed the DC resistivity in this probe limit. It was found that the DC resistivity is suppressed by spontaneous magnetization and shows a  metallic behavior. This is very similar to the strong correlated phenomenon named CMR effect found in the some manganites.

Next we considered the case with full back reaction and solved the full equations of motion numerically.  It was found that the free energy difference between the condense phase and normal phase, $\Delta\Omega$, is zero at critical temperature and always negative when $T<T_c$.  The phase transition is always a second order one  as one increases the strength of the back reaction.

Main calculations are made in $3+1$ dimensions in this paper, but it can be easily extended into the higher dimension case.  In the latter case, the space rotation symmetry will be broken spontaneously
when the phase transition happens. In addition, we can also generalize this model to the case with the Lifshitz symmetry in the bulk  and study  the influences of the Lifshitz dynamical exponent $z$ on the condensate both in the probe limit and in the case with the back reaction. In addition, in all the calculations in  this paper, we assumed that  the solution in the bulk or the phase at the boundary is homogeneous, which is a strong assumption. In fact, inhomogeneous solution may exist even in a model whose Lagrangian has translation symmetry and the boundary conditions are homogeneous (see Ref.~\cite{Ling:2014saa}, for example). In many materials, the inhomogeneous phase can appear spontaneously in a chemical homogeneous materials. Therefore it is of  great interest to study whether the inhomogeneity could appear spontaneously in this model. Although we focused on the ferromagnetic phase transition only in this paper, which has been understood well in condensed matter physics, it offers a  framework rather than only a specific model, which can be regarded as basic starting point to understand more complicated phenomenon involving spontaneous magnetization. As a result, there are various prospects to study and we expect more exciting results could be reported in the future.

\section*{Acknowledgments}
 This work was supported in part by the National Natural Science Foundation of China ( No.11375247 and No.11435006 ), National Natural Science Foundation of China (Grant Nos.11175077) and the Doctoral Program of Higher Education, Ministry of Education, China (Grant
No. 20122136110002)

\appendix
\section{Ginzburg-Landau theory for Ising universality class}
\label{app1}
In this appendix, we will compute the grand thermodynamic potential of the Ising-like Hamiltonian \eqref{Ising2}. We can rewrite it into the form of  $\lambda\phi^4$ theory,
\begin{equation}\label{phi4form}
H=\int d^2x\left[-\frac12(\widetilde{s}\overrightarrow{\nabla}^2\widetilde{s}+m_s^2\widetilde{s}^2)+\frac{\lambda_s\widetilde{s}^4}{4!}\right].
\end{equation}
Turn into the Lagrangian form\footnote{Here we assume the relativistic depression relation.} and go to the case with Euclidian signature, we have the action,
\begin{equation}\label{phi4form}
S_E=\frac12\int_0^\beta d\tau\int d^2x[(\partial_\tau\widetilde{s})^2+(\overrightarrow{\nabla}\widetilde{s})^2-m_s^2\widetilde{s}^2+\frac{\lambda_s\widetilde{s}^4}{12}],
\end{equation}
with $\beta=1/T$, the inverse temperature of the system. We see that the mass dimension of coupling constant $[\lambda_s]=1$, so effective action~\eqref{phi4form} is renormalizable. In order to compute the quantum effective potential, we take the mean field expansion method. We split as usual the field into classical background part and quantum fluctuation  part,
\begin{equation}\label{splitfluc}
\widetilde{s}=\widetilde{s}_{cl}+\eta.
\end{equation}
The quadratic part of $\eta$ in the action is then controlled by a kinetic operator of the form,
\begin{equation}\label{quadrceta}
\frac{\delta^2S_E}{\delta^2\widetilde{s}}[\widetilde{s}_{cl}]=-\partial_\tau^2-\nabla^2+M^2(\widetilde{s}_{cl}),
\end{equation}
where
\begin{equation}\label{Mass1}
M^2(\widetilde{s}_{cl})=-m_s^2+\frac{\lambda_s\widetilde{s}_{cl}^2}2.
\end{equation}
In addition, the Euclidian action contains the linear term of $\eta$, but this term contributes nothing and can be dropped.  And, if we neglect the cubic and quartic order terms of $\eta$, we can obtain  the result in mean field approximation by only taking the tree level of $\eta$ into account. This is a good approximation if $\lambda_s$ and $\lambda_s\widetilde{s}_{cl}$ are small\footnote{In the vicinity of critical temperature, $\lambda_s\widetilde{s}_{cl}$ is always small. So in this region, the cubic  term can always be neglected. However, $\lambda_s$ in general may not be a small quantity. In this case, loop-corrections must be considered.}.
Then it is easy to find the grand thermodynamic potential density  in the mean field approximation as,
\begin{equation}\label{Omegamea}
    \Omega=-\frac{1}{2} m_s^2\widetilde{s}_{cl}^2+\frac{1}{4!}\lambda_s\widetilde{s}_{cl}^4+\int\frac{d^2k}{4\pi^2}[\frac{\omega}2+T\ln(1-e^{-\beta\omega})].
\end{equation}
Here $\omega=\sqrt{k^2+M^2(\widetilde{s}_{cl})}$. The first term in the integral of \eqref{Omegamea} contributes the zero-point energy and can be neglected. Then the second term  in the integral of \eqref{Omegamea} then can be written as,
\begin{equation}\label{Omegamean2}
    \frac{T^3}{2\pi}\int_0^\infty dx\,x\ln(1-e^{-\sqrt{x^2+x_0^2}})],
\end{equation}
where $x_0^2=\beta^2 M^2(\widetilde{s}_{cl})$. The result of \eqref{Omegamean2} does not  admit  a compact expression. We can get a series expression by high temperature expansion, where  $x_0$ is treated as a small quantity,
\begin{equation}\label{Hightem}
\int_0^\infty dx\,x\ln(1-e^{-\sqrt{x^2+x_0^2}})=\sum_{n=0}^{\infty}c_nx_0^{2n},
\end{equation}
where
\begin{equation}\label{valuecs}
c_0=-\zeta(3),~c_1=\gamma_E/2,\cdots.
\end{equation}
Here $\gamma_E=0.5772\cdots$ is the Euler constant and $\zeta(3)=1.202\cdots$ is the value of Riemann-$\zeta$ function.   In this limit we have,
\begin{equation}\label{Omegamean}
    \Omega\approx-\frac T{2\pi}[\zeta(3)T^2+\frac{\gamma_Em_s^2}2]+\frac12(\frac{\lambda_s\gamma_ET}{4\pi}-m_s^2)\widetilde{s}_{sl}^2 +\frac{\lambda_s\widetilde{s}_{sl}^4}{4!},
\end{equation}
from which  we can obtain the expression in~\eqref{Veffs1}.

\section{Semi-analytic calculations near the critical temperature}
\label{app2}
\subsection{Spontaneous magnetization}
In this appendix, we will compute the values of  $N_1$, $a_0$ and $a_1$ appearing in section \ref{prob1}. With these we can get the coefficients in~\eqref{onshellN}  and  compare with the numerical results in  the previous sections.

Let us first compute $N_1$ and $a_1$.  For this we have to first  find  the eigenfunction $\rho_1$, which is the solution of,
\begin{equation}\label{eqrho1}
-\frac{d}{dz}\left[f(z)\frac{d\rho_n}{dz}\right]+q(x)\rho_n=0
\end{equation}
at $T=T_c$ with the conditions,
\begin{equation}\label{cond1}
\rho_1(1)=1,~~~\rho_{1+}=0.
\end{equation}
For convenience, here we do not assume that   $\{\rho_n\}$  form an  unit base. Thus we have,
\begin{equation}\label{N1s}
N_1=\frac1{C_1}\int_{0}^{1}\rho_1dz,~~a_1=\frac1{4C_1^4}\int_{0}^{1} z^8\rho_1^4 dz,
\end{equation}
here $C_1$ is the normalization coefficient and
\begin{equation}\label{eqC}
C_1^2=\langle\rho_1,\rho_1\rangle=\int_{0}^{1} \omega\rho_1^2dz.
\end{equation}

\begin{figure}
\includegraphics[width=0.23\textwidth]{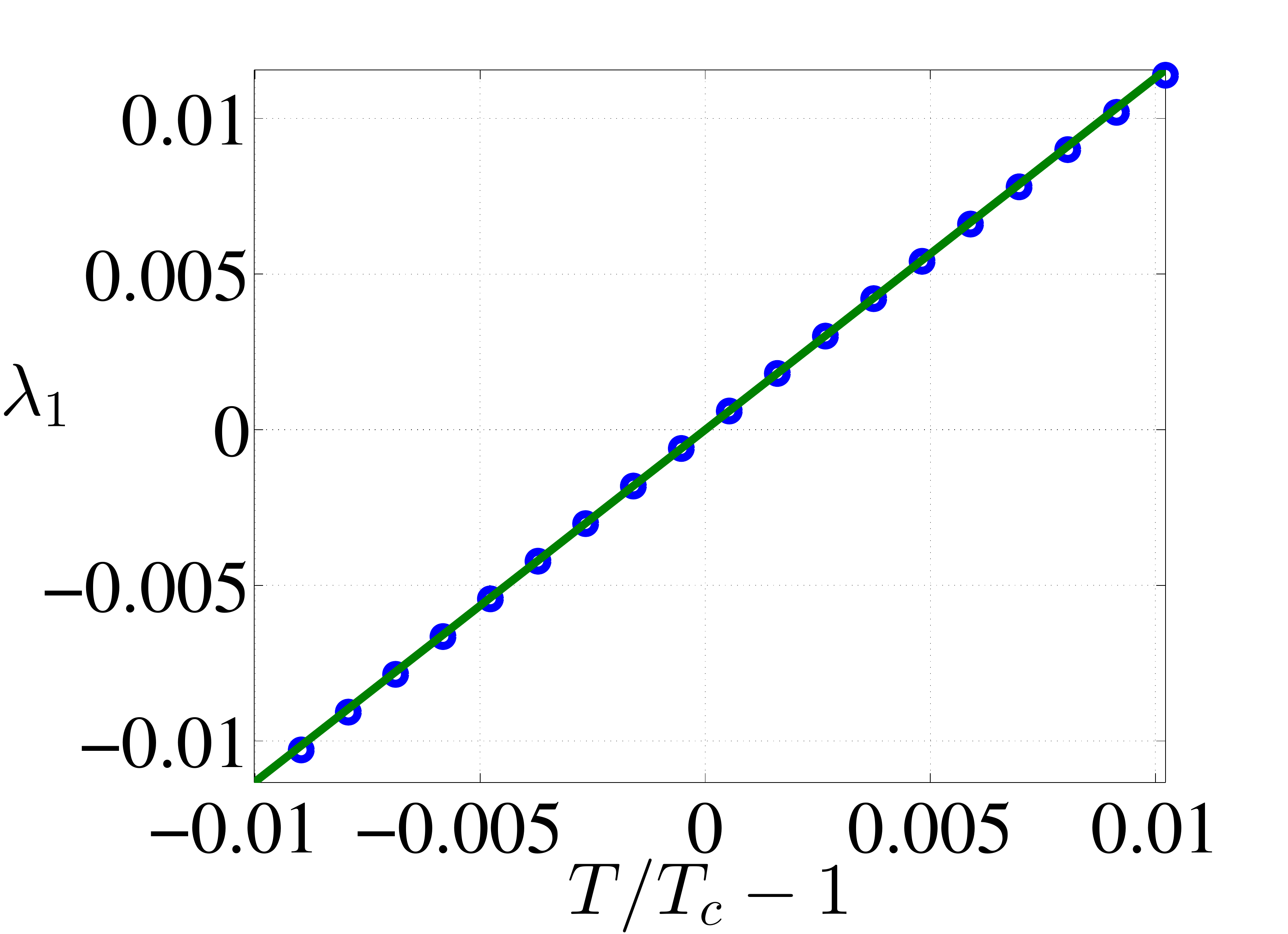}
\includegraphics[width=0.23\textwidth]{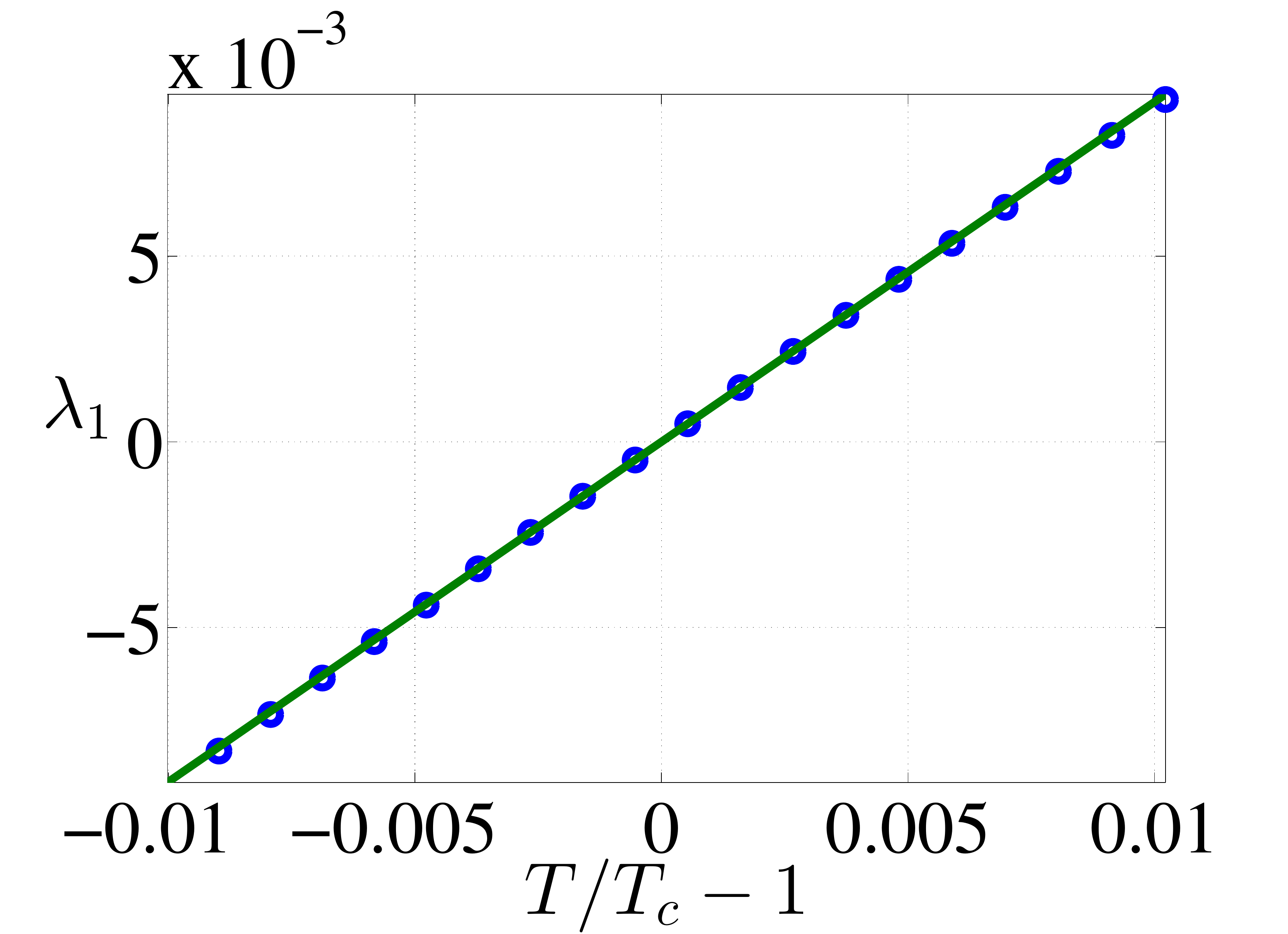}
\caption{The value of $\lambda_1$ with respective to temperature in the case of $m^2=-J=1/8$ and $\omega=z^4$(left) and $\omega=z^3$(right).}
\label{Ta0}
\end{figure}

In order to compute $a_0$, we need to solve equation~\eqref{eqrho1} in the limit  $T\rightarrow T_c^-$.
To clarify that the results are independent of the specific form of weight function, we choose $k=3,4$ as two examples.
Then we fit the relation $\lambda_1=a_0(T/T_c-1)$ to find $a_0$. Figure~\ref{Ta0} shows that $\lambda_1$ and $T/T_c-1$  indeed satisfies a linear relation very well.
Numerical results show that $N_1\simeq2.0412$, $a_1\simeq0.7276$,$a_0\simeq1.1309$ for $k=4$ and $N_1\simeq1.8370$, $a_1\simeq0.4772$,$a_0\simeq0.9155$ for $k=3$.
 We   have,
\begin{equation}\label{onshellN2}
N^2/\mu_c^2=\frac{N_1^2a_0}{-16\widetilde{J}_fa_1\mu_c^2}(1-T/T_c)\simeq a_2(1-T/T_c).
\end{equation}
with $a_2\simeq4.966$ for $k=4$ and $a_2\simeq4.964$ for $k=3$. We  see that different weight functions give different values for $N_1$, $a_1$ and $a_0$,  but the same value of magnetic moment $N$(up to a numerical error). 

The value of $a_0$ can also be obtained directly by  solving ODE~\eqref{SL1}. In the region near the critical temperature, we assum $\lambda_1=a_0(T/T_c-1)$. Note that all quantities in~\eqref{SL1} are the functions of temperature, so take derivative with respect to $T$ and evaluate at $T=T_c$, we get,
\begin{equation}\label{DSL1}
\frac{d\widehat{P}}{dT}\rho_1+\widehat{P}\frac{d\rho_1}{dT}=\frac{a_0}{T_c}\rho_1.
\end{equation}
Here $\rho_1$ is the eigenfunction of~\eqref{eqrho1}.  Now treat $\rho_T=\frac{d\rho_1}{dT}$ as an unknown function to be solved, then the task to find $a_0$ becomes to solve  a non-homogenous eigenvalue problem,
\begin{equation}\label{DSL2}
\widehat{P}\rho_T=\left[\frac{a_0}{T_c}-\frac{d\widehat{P}}{dT}\right]\rho_1.
\end{equation}
At the AdS boundary, $\rho_T$ has the same asymptotic behavior as~\eqref{asym1},  thus  we can impose the boundary conditions as
\begin{equation}\label{bound}
|\rho_T(1)|<\infty,~\rho_{T+}=0.
\end{equation}
We find that $\rho_T\in D$.   We  then  use the  basis  $\{\rho_n\}$ to expand $\rho_T$, i.e.,
\begin{equation}\label{expanda1}
\rho_T=\sum_{n=1}^\infty \frac{d_n}{C_n}\rho_n.
\end{equation}
Here $C_n$ are  the modules of $\rho_n$. Using the fact $\lambda_1=0$ at $T=T_c$ and
\begin{equation}\label{expanda2}
\begin{split}
\langle C_1^{-1}\rho_1,\widehat{P}\rho_T\rangle&=\sum_{n=1}^\infty d_n\langle C_1^{-1}\rho_1,C_n^{-1}\widehat{P}\rho_n\rangle\\
&=\sum_{n=1}^\infty d_n\lambda_n\delta_{1n}=d_1\lambda_1=0.
\end{split}
\end{equation}
we have,
\begin{equation}\label{DSL3}
\begin{split}
\langle \rho_1,\left[\frac{a_0}{T_c}-\frac{d\widehat{P}}{dT}\right]\rho_1\rangle&=\frac{a_0C_1^2}{T_c}-\int_{0}^{1} \omega\rho_1\frac{d\widehat{P}}{dT}\rho_1 dz\\
&=\frac{a_0C_1^2}{T_c}-\int_{0}^{1}\rho_1\frac{d\widehat{L}}{dT}\rho_1 dz=0.
\end{split}
\end{equation}

Furthermore  we get,
\begin{equation}\label{DSL4a}
a_0=\frac{T_c}{C_1^2}\int_{0}^{1} dz\rho_1\frac{d\widehat{L}}{dT}\rho_1.
\end{equation}
It is very useful to find its equivalent form in the case  with fixed $r_h=1$, since it is convenient when we perform numerical computation. If we fix $r_h=1$, the shooting parameter is chemical potential $\mu$. The relation between temperature in grand canonical ensemble and chemical potential is given by,
\begin{equation}\label{relTmu}
T=\frac{3-\mu^2}{4\pi\mu}.
\end{equation}
Thus the expression \eqref{DSL4a} can be rewritten as
\begin{equation}\label{DSL4}
\begin{split}
a_0&=\frac{T_c}{\mu_c^2C_1^2}\frac{d\mu^2}{dT}\int_{0}^{1} dz\rho_1\frac{d\widehat{L}}{d\mu^2}\rho_1 \\
&=-\frac{8\mu_c^3\pi T_c}{3C_1^2}\int_{0}^{1} dz\rho_1\left.\frac{d\widehat{L}}{d\mu^2}\rho_1\right|_{\mu=\mu_c} .
\end{split}
\end{equation}
Here $\mu_c$ is the critical chemical potential when we fix $r_h=1$ and $T_c=\frac{3-\mu_c^2}{4\pi\mu_c}$ is the critical temperature in grand canonical ensemble. Combining~\eqref{N1s} and~\eqref{DSL4}, we have,
\begin{equation}\label{onshellN2}
N^2/\mu_c^2\lambda^4=\frac{\pi \mu_cT_c(\int_{0}^{1}\rho_1dz)^2\int_{0}^{1} dz\rho_1\frac{d\widehat{L}}{d(\mu^2)}\rho_1}{24\widetilde{J}_f\int_{0}^{1} dzz^4\rho_1^4}(1-T/T_c).
\end{equation}
We can see that it is determined by the equation~\eqref{eqrho1}, but  independent of the weight function!  The expression~\eqref{DSL4} depends on the weight function, because it depends on  $C_1$. In order to check the formula \eqref{DSL4},  let us compute the values of $a_0$ in the cases of $k=4$ and $k=3$. The results gives  $1.1286$ and 0.9140, respectively. We see that, up to numerical errors, they are the same as what we have obtained by fitting the curve in~figure~\ref{Ta0}. In addition, the expression ~\eqref{onshellN2} gives $a_2=4.9560$, which is very close to what we obtained in the numerical calculation.

\subsection{Susceptibility and hysteresis loop}
When $B\neq0$, the susceptibility for $T>T_c$ is defined as,
\begin{equation}\label{sus1}
\chi=\lim_{B\rightarrow0}\left(\frac{\partial N}{\partial B}\right)_T.
\end{equation}
In the case with  $T>T_c$ and $B\rightarrow0$, we can neglect the non-linear term, i.e., setting $\widetilde{J}_f=0$.  The solution of equation~\eqref{rhop}  can be expressed as,
\begin{equation}
\rho=\sum_{n=1}^\infty c_n\rho_n-\frac{B}{m^2},
\end{equation}
Taking into account the equation~\eqref{rhop} with $\widetilde{J}_f=0$, we have
\begin{equation}\label{rhop2}
\begin{split}
0&=\widehat{L}\rho+B=\sum_{l=1}^\infty c_l\widehat{L}\rho_l-\frac{4 BJ \mu^2z^4}{m^6}\\
&=\sum_{l=1}^\infty c_lC_l^{-1}\lambda_l\omega\rho_l-\frac{4 BJ \mu^2z^4}{m^6}.
\end{split}
\end{equation}
Multiplying a factor  $\rho_n/C_n$ and integrating the above equation from 0 to 1, we can obtain
\begin{equation}\label{expandB1}
c_n=\frac{B\gamma_n}{\lambda_n},~~\text{with}~\gamma_n=\int_{0}^{1}\frac{4J \mu^2z^4}{C_n m^6}\rho_ndz.
\end{equation}
Thus we can  get the magnetic moment density as
\begin{equation}\label{meg1}
N/\lambda^2=\frac{B}{2m^2}-B\sum_{n=1}^\infty \frac{\gamma_nN_n}{2\lambda_n},
\end{equation}
and the magnetic susceptibility
\begin{equation}\label{chi1}
\chi/\lambda^2=\frac1{2m^2}-\sum_{n=1}^\infty \frac{\gamma_nN_n}{2\lambda_n}.
\end{equation}
When $T\rightarrow T_c^+$, we have $\lambda_1=a_0(T/T_c-1)\rightarrow0^+$.  Thus  $\chi$ is  dominated by the first term in the summation of~\eqref{chi1} and its inverse can be expressed as

\begin{equation}\label{chi2}
\lambda^2\chi^{-1}/\mu_c=-\frac{2a_0}{\mu_c\gamma_1N_1}(T/T_c-1),~~\text{as}~T\rightarrow T_c^+.
\end{equation}
In the case of $m^2=-J=1/8$, we have  $\lambda^2\chi^{-1}/\mu_c\simeq4.0520(T/T_c-1)$, which is very close to our numerical result   $\lambda^2\chi^{-1}/\mu_c\simeq4.0499(T/T_c-1)$ given in the
numerical calculation.

\begin{figure}[h!]
\begin{center}
\includegraphics[width=0.35\textwidth]{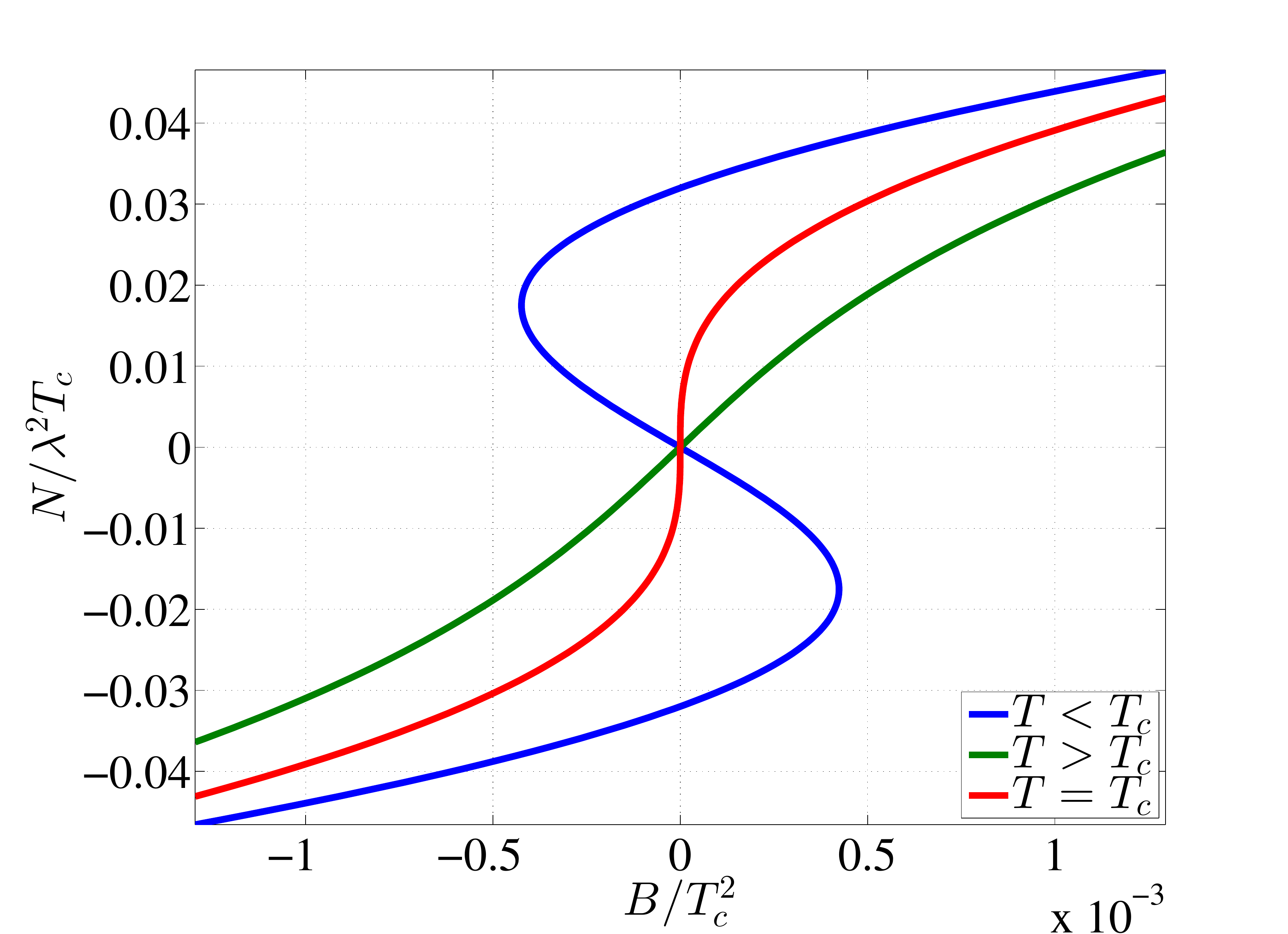}
\caption{The relation between  magnetic moment density $N$ and external magnetic field $B$  in the cases of  $T=1.05T_c, T=0.9T_c$ and $T=T_c$, respectively.}
\label{relBN}
\end{center}
\end{figure}

Next  let us move to the case  with  $B\neq0$.  In this case  from  \eqref{rhop} we have,
\begin{equation}\label{eqB1}
\int_{0}^{1}\rho_n(\omega\widehat{P}\rho+B-\widetilde{J}_f\rho^3z^8)dz=0.
\end{equation}
Consider \eqref{exprho1}, we can rewrite it to
\begin{equation}\label{eqBN}
\int_{0}^{1}\rho_n(\omega\widehat{P}\widetilde{\rho}+B[1-\frac{q(z)}{m^2}]-\widetilde{J}_f\rho^3z^8)dz=0.
\end{equation}
Using the expansion expression~\eqref{exprho1}, we have,
\begin{equation}\label{eqBN2}
c_nC_n^2\lambda_n+B\gamma_n-\int_{0}^{1}\rho_n\widetilde{J}_f\rho^3z^8dz=0,~~n=1,2,\cdots.
\end{equation}
For convenience, we assume that $\{\rho_n\}$ is an unit base, i.e., $C_n$=1. The equations~\eqref{eqBN2} are equivalent to \eqref{rhop} if we take all the terms in \eqref{exprho1} into account. In the case of $T\rightarrow T_c^-$, assume that  the first term in \eqref{exprho1} dominates only, i.e., $|c_1|\gg c_n$ for $n\geq2$ in \eqref{coeffN},  we have
\begin{equation}
\label{eqN2}
N/\lambda^2=\frac{B}{2m^2} -c_1N_1/2.
\end{equation}
Taking $n=1$ in \eqref{eqBN2}, we have,
\begin{equation}\label{eqBN3}
c_1\lambda_1+B\gamma_1-c_1^3\widetilde{J}_f\int_{0}^{1}\rho_1^4z^8dz=c_1\lambda_1+B\gamma_1-4c_1^3\widetilde{J}_fa_1=0.
\end{equation}
For a given temperature $T\rightarrow T_c$, we can combine~\eqref{eqN2} and~\eqref{eqBN3} to obtain a relation between  external magnetic field $B$ and magnetic moment density $N$. Figure~\ref{relBN} shows the results with $T=1.05T_c, T=0.9T_c$ and $T=T_c$, respectively,  in the case of $m^2=-J=1/8$. We  see that it is very similar to what we have obtained in the previous work~\cite{29}.

Finally, let us notice that in GL theory, the equation for magnetic moment density is,
\begin{equation}\label{GLNBT}
A_1(T-T_c)N+A_2N^3-B=0
\end{equation}
with two positive coefficients $A_1$ and $A_2$.  However,  it is easy to see that the equation for $N$ in our model is different from the usual form~\eqref{GLNBT} from the GL theory, which can be
obtained by combining \eqref{eqN2} and~\eqref{eqBN3} to eliminate $c_1$. Namely, alhought our model gives the similar results  near the  critical temperature in GL theory, there exist  some differences
between the holographic model and the GL theory  even in the region near the critical temperature.

\end{document}